\documentclass[11pt,a4paper]{article}
\usepackage[colorlinks,bookmarksopen,bookmarksnumbered,citecolor=red,urlcolor=blue]{hyperref}

\usepackage{multibib}
\usepackage[T1]{fontenc}         
\usepackage[latin1]{inputenc}
\usepackage{graphicx}
\usepackage{amsmath,amssymb}
\usepackage{bm}
\usepackage{tikz}
\usepackage{helvet}
\usepackage{courier}
\usepackage{makeidx}
\usepackage{footmisc}
\usepackage{type1cm}         
\usepackage{textcomp}
\usepackage{subfig}
\usepackage{yhmath}
\usepackage{esdiff}
\usepackage{mathtools}
\usepackage{color}
\usepackage{enumitem,xcolor}    
\usepackage{pdfpages}

\DeclareCaptionLabelSeparator{colon}{. }
\usepackage[labelfont={bf},labelsep={colon}]{caption}

\makeatletter
\DeclareRobustCommand\dalemb{\mathpalette\inner@dalemb{}}
\def\inner@dalemb#1{%
  \add@dalemb#1{03}%
  \add@dalemb#1{06}%
  \square
}
\def\add@dalemb#1#2{%
  \sbox0{\scalebox{1.#2}{$#1\square$}}%
  \rlap{\lower0.#2\ht0\box0}%
}
\makeatother
\DeclareMathOperator{\dalembert}{\Box}

\makeindex

\DeclareMathAlphabet\mathbfcal{OMS}{cmsy}{b}{n}

\begin{document}
\thispagestyle{empty}
\bibliographystyle{elsarticle-num} 



\begin{center}
\textcolor{blue}{ \Large  \bf Application of discrete mechanics model to jump conditions in two-phase flows} \\
\vspace{3.mm}
{\bf Jean-Paul Caltagirone } \\
\vspace{3.mm}
{ \small Universit{\'e} de Bordeaux  \\
   Institut de M{\'e}canique et d'Ing{\'e}ni{\'e}rie \\
   D{\'e}partement TREFLE, UMR CNRS  n\textsuperscript{o}5295\\
  16 Avenue Pey-Berland, 33607 Pessac Cedex  \\
\textcolor{blue}{\texttt{ calta@ipb.fr }  } }
\end{center}

\textcolor{blue}{\bf Abstract}

Discrete mechanics is presented as an alternative to the equations of fluid mechanics, in particular to the Navier-Stokes equation. The derivation of the discrete equation of motion is built from the intuitions of Galileo, the principles of Galilean equivalence and relativity. Other more recent concepts such as the equivalence between mass and energy and the Helmholtz-Hodge decomposition complete the formal framework used to write a fundamental law of motion such as the conservation of accelerations, the intrinsic acceleration of the material medium, and the sum of the accelerations applied to it. The two scalar and vector potentials of the acceleration resulting from the decomposition into two contributions, to curl-free and to divergence-free, represent the energies per unit of mass of compression and shear.

The solutions obtained by the incompressible Navier-Stokes equation and the discrete equation of motion are the same, with constant physical properties. This new formulation of the equation of motion makes it possible to significantly modify the treatment of surface discontinuities, thanks to the intrinsic properties established from the outset for a discrete geometrical description directly linked to the decomposition of acceleration. The treatment of the jump conditions of density, viscosity and capillary pressure is explained in order to understand the two-phase flows. The choice of the examples retained, mainly of the exact solutions of the continuous equations, serves to show that the treatment of the conditions of jumps does not affect the precision of the method of resolution.

\vspace{2.mm}

\textcolor{blue}{\bf Keywords}

Discrete Mechanics; Hodge-Helmholtz Decomposition; Navier-Stokes equation; Two-Phase Flows; Mimetic Methods; Discrete Exterior Calculus

\vspace{-3.mm}
\begin{verbatim}
_____________________________________________________________________
\end{verbatim}
\vspace{-2.mm}
This article may be downloaded for personal use only. Any other use requires prior permission of the author and Elsevier Inc. 
\vspace{1.mm}

J-P Caltagirone, Application of discrete mechanics model to jump conditions in two-phase flows, Journal  of Computational Physics, 2021,  https://doi.org/10.1016/j.jcp.2021.110151.
\vspace{-6.mm}
\begin{verbatim}
_____________________________________________________________________
\end{verbatim}

\vspace{-5.mm}

\normalsize

\textcolor{blue}{\section{Introduction} }

The Navier-Stokes equation in its various formulations forms the indisputable base of fluid mechanics, both from a physical point of view and from its ability to provide reliable predictions through its numerical solution. Discrete mechanics does not call into question the validity of these equations, but can be presented as an alternative whose relevance can only be evaluated in the light of a long test. The case of two-phase flows is one of the relatively recent fields covered by computational fluid dynamics. 

Two-phase incompressible flows have particular specificities, such as the presence of interfaces formed by several fluids, generating capillary effects or even very large variations in physical properties. One of the problems generated by variable interface flows over time is that of advection. There are many ways of transporting these surfaces, such as Volume Of Fluid, Level-Set, Front-Tracking, Arbitrary Lagrangian Eulerian, or combinations of these methods. Many articles highlight the defects and qualities of each \cite{Sca99}, \cite{Pro07}. The problems of solid inclusion flow also feature in the very abundant literature on theoretical formulations and their implementation.
Numerical treatment of discontinuities has naturally been addressed for highly compressible flows, particularly by Fedkiw \cite{Fed99} and many other authors using the Ghost Fluid method \cite{Wan06}. The case of discontinuities related to scalar equations is also addressed by this same technique \cite{Gui15}, \cite{Liu17}. The treatment of discontinuities for two-phase flows makes extensive use of this method \cite{Trontin20126990}.
Other popular numerical approaches consist in using a cut cell method on a Cartesian mesh for incompressible flows \cite{Tuk00} or as an alternative to traditional boundary fitted grid methods for compressible flows with shock-waves \cite{Ing03}.

The range of difficulties encountered during two-phase-flows simulations is reflected in various errors:
(i) advection of interfaces, (ii) calculation of curvatures, (iii) location of the interface in the mesh, (iv) assignment of physical properties by interpolations on the stencil, and (v) implementation of jumps in the numerical formulation. The objective is to present a formulation and numerical treatment of jumps in physical properties, density and viscosity, as well as capillary effects directly associated with two-phase flows.

The formal framework is that of discrete mechanics \cite{Cal19a}, which can be presented as an alternative physical model of the Navier-Stokes equations. The associated discrete formulation abandons the notion of continuous medium in favor of an equation on bases of differential geometry that are slightly comparable to the DEC (Discrete Exterior Calculus) methods \cite{Des05} or mimetic methods \cite{Pal17}. Each physical effect of the equation is represented by a solenoidal term and an irrotational one following a Hodge-Helmholtz decomposition. The jumps in the physical properties and those resulting from the capillary effects are also represented in the same way. Unlike other techniques, jumps are located within a single mesh on the one hand and are completely implicit within the equation of motion. In particular, no additional ghost point is needed.

To present the advantages of the theoretical formulation, the solutions of degree lower or equal to two are used to validate it. As the methodology of discrete mechanics is of order two in space, it allows precise simulations to be carried out to the machine error, whatever the regular meshes adopted and the number of degrees of freedom. A second-order accurate method applied to a complex problem can cover multiple errors, the point here being to verify that the discrete formulation is free of all {\it artifacts}.

\textcolor{blue}{\section{Discrete formulation} }

\textcolor{blue}{\subsection{Framework of discrete mechanics} }

\textcolor{blue}{\subsubsection{Primal and dual geometric topologies} }

This section defines the formal framework for deriving the equation of motion. Notions substantially different from those of continuous media must be specified. First of all, the Galilean or classical inertial frame of reference is abandoned and replaced by a local frame of reference where the interactions are of cause and effect, i.e. the information propagates with a celerity $c$ from one local frame of reference to another. It is then no longer possible to change the frame of reference in the usual sense of the term because the celerity is not a constant quantity, it varies according to the medium. The concept of continuous medium disappears and the physical quantities, characteristics and variables are no longer defined at a point; they are located on different positions of the local frame of reference, the choice of which is dictated by the coherence of the physical effects described.

\begin{figure}[!ht]
\begin{center}
\includegraphics[width=5.cm]{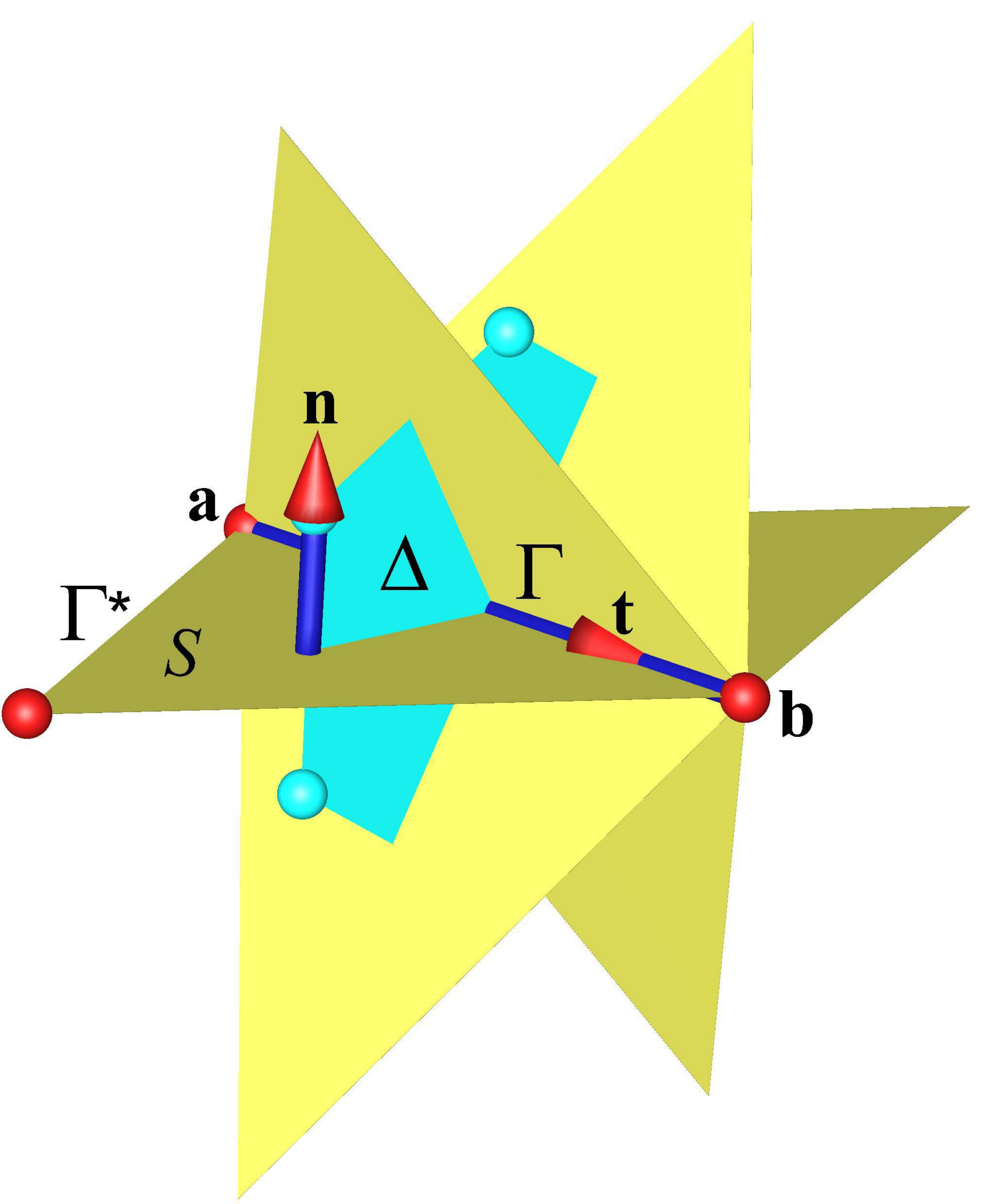}
\caption{\it Discrete geometric topology: a set of primitive planar facets $\mathcal S$  are associated with the segment $\Gamma$ of unit vector $\mathbf t$ whose ends $a$ and $b$ are distant by a length $d$. Each facet is defined by an contour $\Gamma^*$, a collection of 3 segments $\Gamma$, is oriented according to the normal $\mathbf n$ such that $\mathbf n \cdot \mathbf t = 0$; the dual surface $\Delta$ connecting the centroids of the cells is also flat. }
\label{primdual}
\end{center}
\end{figure}

Similarly, a vector is no longer defined by its Cartesian components in a global frame of reference $ (x, y, z) $. If $ \mathbf W $ is a vector of $\mathcal R^3$, discrete mechanics defines the component of $ \mathbf W $ both as a scalar on a segment $ \Gamma $ directed by the unit vector $ \mathbf t $ and as a vector $\mathbf V = (\mathbf W \cdot \mathbf t) \: \mathbf t$. Only the quantity $ \mathbf V $ is considered in the formulation; vector $ \mathbf W $ will remain unused even if it can be reconstructed from the components in a three-dimensional space, but also on each of the facets of the primal geometry. Thus, the gradient operator of a scalar is not the vector of space in the usual sense but its projection on the segment of unit vector $\mathbf t$; it will be the same for the other differential operators used. 
The notion of gradient of a vector is non-existent, like all tensors of order equal to or greater than two in classical mechanics. The concept of tensor created for and by mechanics comes directly from the observation of media having characteristics which depend on the direction considered, wood or quartz for example. This natural 18th-century idea to transpose the directional characteristics of environments to the modeling of mechanical effects can be discussed and modified.
In fact, the adoption of a $(x, y, z)$ three-dimensional space description three centuries ago persists in present-day mechanics and physics, for example in the theory of relativity with a four-vector concept. The abandonment of the concept of continuous medium entails that of derivation, integration and analysis. Discrete mechanics reconstructs a discrete physical model based on simple operators applied to scalars, the vertices and the barycenters of the facets associated with the normals of the primal geometry.
Figure (\ref{primdual}) specifies the elementary discrete geometry, constituted by a rectilinear segment $\Gamma$ of ends $a$ and $b$ and of planar facets formed by a collection $\Gamma^*$ of segments delimiting polygonal facets $\mathcal S$ having a common side with the segment $\Gamma$.

The unit vector $ \mathbf n $ located at the barycenter of each facet is orthogonal to that carried by the segment $ \Gamma $, the vector $ \mathbf t $, $ \mathbf n \cdot \mathbf t = 0 $. The surface $ \Delta $ is not, in the general case, necessarily planar if the polygons delimiting the facets are not regular. This point will be discussed during the processing of the jump conditions but, for this step establishing the physical model, this will be the case. Thus the facets will be regular polygonal, equilateral triangles, quadrangles, hexagons, etc. The physical domain will be tessellated by the pattern represented by figure (\ref{primdual}), the primal mesh.

A primary physical meaning must be attributed to the elementary stencil of figure (\ref{primdual}); it is more easily understood in elementary electromagnetism: an electric current in a rectilinear conductor produces a magnetic field in its vicinity and conversely a magnetic field produces an electric current. Thus a flow represented by a velocity on $\Gamma$ produces a field of rotation, the curl of velocity $ \nabla \times \mathbf V $, an axial vector carried by the unit vector $\mathbf n$. The variation of the velocity on the segment $\Gamma$ is thus produced by two fundamental actions: a direct action generated by a potential difference between $a$ and $b$ and another action induced by the circulation of the axial vector on the contour of $\Delta$. These two contributions of the acceleration have no interactions between them because they are associated with orthogonal quantities. They can only exchange information, energy for example, if the phenomenon is dynamic; J.C. Maxwell \cite{Max65} expressed this dynamic duality very well in the pathway which led him to the equations which bear his name.

The acceleration denoted $\bm \gamma$ is likewise a component of the acceleration vector, knowledge of which is not useful for physical modeling or the derivation of the equation of motion; like velocity, it will be constant on the segment $\Gamma$. In discrete mechanics it has the status of absolute quantity and can be measured without any reference to the external environment; the other quantities - velocity, pressure, energy, potentials, etc. - are relative and defined up to constants which must be filtered by the discrete operators in order to ensure the invariance of the equation of motion.

Space and time are linked by the local celerity of the propagation phenomenon considered (swell, sound, light) through the notion of discrete horizon $ dh $, such that $dh = c \: dt$ where $dt$ is the  elapsed time between two states of mechanical equilibrium. The discrete horizon is of the order of magnitude of the length $d = [a, b]$ of the segment $\Gamma$. The state of mechanical equilibrium is itself defined as the exact satisfaction of the equation of motion; $t^o$ is the reference state and $t = t^o + dt$ is the current state. The equation of motion must allow the calculation of the set of variables at time $t$ from that at time $t^o$ in an incremental time process.

\textcolor{blue}{\subsubsection{Physical model} }

The physical model is developed from the following principles and postulates:
\vspace {2.mm}
\begin{itemize} [label = \textcolor {blue} {\textbullet}]
\item Galileo's remarkable intuition on the principle of weak equivalence between gravitational and inertial effects of mass. The equivalence between ``the grave mass'' and ``the inertial mass'' \cite{Wil18} is now verified at less than one part on $10^{15}$, hence its status as a principle;
\item the Galilean principle of velocity relativity, which should lead to an invariant physical model for uniform motion;
\item the principle of equivalence of energy and mass formulated by A. Einstein, resulting from work on the theory of relativity;
\item the Helmholtz-Hodge decomposition where any vector can decompose into a divergence-free component and another curl-free component.
\end{itemize}

The principle of equivalence stems from the experiments of the beginning of the 16th century, in particular that of Galileo where two different masses fall with the same acceleration and the same velocity; Galileo himself attributes this observation to the fact that the mass related to inertia is equal to the mass attached to gravity. The weak equivalence principle (WEP)  states that the inertial mass and the gravitational mass are equal. Whatever the forms adopted thereafter, i.e. the equivalence principle of general relativity, strong equivalence principle etc., presented as a local equivalence between gravitation and acceleration, mass is always present there. Singularly, while the phenomenon does not depend on mass, this concept still persists today where mass is most often attached to acceleration or velocity. Discrete mechanics returns to the interpretation of this principle based on the original observation, with the equality of the gravitational acceleration and the intrinsic inertial acceleration thus expressing the instantaneous mechanical equilibrium of a body.

The fundamental law of dynamics or Newton's second law translates, in its modern version, the equality between the variation of the momentum of a body and the sum of the external forces, i.e. $ m \: \bm \gamma = \mathbf F $. In this presentation, as the velocities are much lower than the celerity of light, the moving mass $m$ will be assimilated to the rest mass $m_0$. When the force is linked to gravity, the law of dynamics becomes $m \: \bm \gamma = m \: \mathbf g $ or $ \bm \gamma = \mathbf g$, an equality between accelerations. The laws of mechanics are still based today on the conservation of momentum $ \mathbf q = m \: \mathbf V $; in fluid mechanics the Navier-Stokes equation is a law of conservation of momentum. In physics the theory of relativity also considers momentum in the quadrivector formulation. The upkeep of mass or density in the equation of motion was due, until the 20th century, to the legitimately predominant role of gravity. What is true for gravity could be extended to other accelerations. In discrete mechanics the fundamental law of dynamics translates that the intrinsic acceleration of a material medium on a segment is equal to the sum of the accelerations which are applied to it:
\begin{eqnarray}
\displaystyle{ \bm \gamma = \bm h } 
\label{galilee}
\end{eqnarray}
where $\bm h$ is the sum of the accelerations, those of the effects of compression and shear but also all the other potential source terms: gravitation, capillary acceleration, etc.

It is essential for the equation of motion to be invariant with respect to uniform motions and, of course, the uniform translational motion at constant velocity; this is Galilean relativity. The invariance must be extended to the uniform rotational motion at constant velocity $\mathbf \Omega$ such that $\mathbf V = \mathbf \Omega \times \mathbf r$. In discrete mechanics the rotational invariance is structurally assured \cite{Cal20c}. Noether's theorem \cite{Noe11} establishes the link between invariances and the conservation of certain quantities. The rotational invariance leads to the conservation of angular momentum.

The equivalence between mass and energy, one of the consequences of the special theory of relativity carried by the law $e = m \: c^2$, induces a simplistic interpretation which consists in asserting that the mass is equal to energy; however, the velocity $\bm v$ of a particle or a material medium is not always equal to the celerity of the wave, whatever its nature (swell, sound, light). The energy per unit of mass is neither equal to $\bm v^2$ nor equal to $c^2$, the equation of motion will give its expression. However, it is certain that the two concepts are concurrent and it is possible to conserve mass or to conserve energy, but the conservation of both is redundant. In all the units of physics, mass only intervenes in the order $1$, $0$ or $-1$ and it is therefore possible to define all these quantities per unit of mass. Discrete mechanics establishes that the equation of motion is both a law of conservation of acceleration and a law of conservation of energy.

Finally, the Helmholtz-Hodge decomposition commonly used in mathematics, for example to project a field onto a space with zero divergence for the \cite{Gue06} projection methods, has only a marginal role in physics. The Helmholtz-Hodge decomposition makes it possible to write a vector in a divergence-free component and another curl-free one, but decomposing the velocity vector in physics involves the relativity of this quantity, which results in the presence of a third harmonic component, both divergence-free and curl-free. The Helmholtz-Hodge decomposition of acceleration is of an entirely different nature because the uniform translational and rotational motions are filtered out by the discrete equation of motion. The law of dynamics thus becomes:
\begin{eqnarray}
\displaystyle{ \bm \gamma = - \nabla \phi + \nabla \times \bm \psi } 
\label{newton}
\end{eqnarray}
where $\phi$ is the scalar potential and $ \bm \psi $ is the vector potential, these are the current potentials (at the instant $ t $) of the acceleration; they are expressed in $ m ^ 2 \: s^{-2}$, the unit of energy per unit of mass. The first term of the right-hand side is an acceleration produced by a direct action due to a difference in scalar potential between $a$ and $b$, while the second is an acceleration due to an induced action generated by the circulation of the vector $\bm \psi$ along the contour of the dual surface $\Delta$.

The derivation of the equation of motion is performed on the segment $\Gamma$ where the components of the three terms of the law (\ref{newton}) are expressed where they are constant.
\begin{eqnarray}
\displaystyle{  \int_{\Gamma} \bm \gamma \cdot \mathbf t \: dl = - \int_{\Gamma} \nabla  \phi  \cdot \mathbf t \:  dl + \int_{\Gamma} \nabla \times \bm  \psi \cdot \mathbf t \:  dl  } 
\label{bilanlin}
\end{eqnarray}

The length $d$ of the segment $\Gamma$ is chosen according to the problem considered and is as small as necessary, but can in no case be reduced to zero; its orientation will thus be preserved. The integration of an acceleration on a segment corresponds to an energy, and that of the intrinsic acceleration is the sum of the compression and rotation energies:
\begin{eqnarray}
\displaystyle{  \Phi_b - \Phi_a = \int_a^b \bm \gamma \cdot \mathbf t \: dl   } 
\label{energy}
\end{eqnarray}
where $\Phi$ is total energy per unit mass.

\textcolor{blue}{\subsubsection{A continuous vision of motion} }

This section is dedicated to the interpretation of discrete vector fields, considering the line integrals along the oriented segments of the geometric topology of figure (\ref{primdual}). The objective is to deduce a local shape of the continuous medium type. The notations are particularly confusing for the primal and dual curls and do not allow one to distinguish primal curl $\nabla \times \mathbf W$, an exact operation, from dual curl $\nabla \times \bm \psi$, an approximate operation. The geometrical reduction towards a local equation is carried out differently, however, by considering a homothetic transformation preserving the angles, in particular. However, this continuous version has the advantage of allowing a direct comparison with the equation of motion of continuum mechanics. The discrete version will be established in section 2.2.

The discrete equation of motion is thus presented as a law of conservation of acceleration but also of mechanical energy; acceleration $ \bm \gamma $, a component of the acceleration vector, is the material derivative of the velocity component on $\Gamma$:
\begin{eqnarray}
\displaystyle{  \frac{d \mathbf V}{d t}  = -  \nabla  \left( \phi^o + d \phi \right) +  \nabla \times \left( \bm  \psi^o + d \bm \psi \right)  } 
\label{loidynhh}
\end{eqnarray}

The law (\ref{newton}) shows the current potentials while at instant $t^o$ only the potentials $\phi^o$ and $\bm \psi^o$ are known; they are called retarded potentials, like those of Li{\'e}nard-Wichert in electromagnetism \cite{Lie98}. During the time period $dt$, the potentials vary in quantities $d \phi$ and $d \bm \psi$ and must be modeled from the longitudinal and transverse celerities $c_l$ and $c_t$, which are different for an elastic medium but equal $ c_l = c_t = c $ for a fluid, where it is commonly called the speed of sound. The physical modeling of $d \phi$ and $d \bm \psi$ are described previously \cite{Cal15, Cal19a}.

The modeling of the variations of the potentials, $d \phi$ and $d \bm \psi$, can be explained simply by considering that these quantities are energies per unit of mass which evolve during a mechanical transformation. The first corresponds to an elementary longitudinal compression of a medium of celerity $c_l$, where the variation of energy is written $d \phi = c_l^2 \: \nabla \cdot \mathbf U$ where $\mathbf U$ is the displacement in the direction considered. The second variation, that of the shear energy, orthogonal to the first, is $d \bm \psi = c_t^2 \: \nabla \times \mathbf U$; the latter is polarizable in a plane orthogonal to direction $\mathbf n$.

Intrinsic acceleration $\bm \gamma$ is then the sum of two contributions where the first, known as direct, is due to the difference in scalar potential between the two ends of segment $\Gamma$. The second is an induced effect of the first; indeed, the velocities carried by contour $\Gamma^*$ create a curl carried by normal $\mathbf n$ to facet $\mathcal S$; this axial vector $\bm \psi$, associated with that of all the other facets having segment $\Gamma$ in common, makes it possible to calculate a circulation on the contour of the dual surface $\Delta$ passing through the barycenters of the facets. This dual curl is carried by segment $\Gamma$, thus defining a second contribution to the intrinsic acceleration $\bm \gamma$:
\begin{eqnarray}
\displaystyle{  \int_{\Gamma} \bm \gamma \cdot \mathbf t \: dl = - \int_{\Gamma} \nabla \left( \phi^o - \: dt \: c_l^2 \: \nabla \cdot \mathbf V \right) \cdot \mathbf t \:  dl + \int_{\Gamma} \nabla \times \left(\bm  \psi^o - dt \: c_t^2 \: \nabla \times \mathbf V \right) \cdot \mathbf t \:  dl  } 
\label{disurG2}
\end{eqnarray}

The second contribution is orthogonal to the first and they cannot combine, but simply overlap; exchanges between the two contributions are only possible if acceleration $\bm \gamma $ is not zero. This phenomenon is in line with J.C. Maxwell's idea on the dynamic role of electromagnetic interactions \cite{Max65}. Equation (\ref{loidynhh}) is a formal Helmholtz-Hodge decomposition of acceleration. The third harmonic term at the same time with divergence-free and curl-free does not exist here because it corresponds to the uniform translational and rotational movements immediately eliminated by the operators in accordance with the principle of relativity. The intrinsic acceleration $\bm \gamma = d \mathbf V / d t$ has a particular status: it is the only quantity that can be considered as absolute. As we perceive that the direct actions represented by $\nabla \phi$ and induced actions given by $\nabla \times \bm \psi$ are entangled, in several dimensions of space one does not exist without the other -- one is the dual of the other. The classical interactions of electromagnetism between direct currents and induced currents are of exactly the same nature in mechanics. 
The variation due to an acceleration of one of the effects implies the variation of the other. The modification of the state of the system is not instantaneous, it respects the principle of causality, it is conditioned by the celerity of the medium.
The formulation has a certain number of properties, in particular the global and local orthogonality of the terms in gradient and in dual curl. The discrete operators mimic some properties of the continuous operators $\nabla \times (\nabla \phi) = 0$ and $\nabla \cdot (\nabla \times \bm \psi) = 0$, irrespective of the polygonal geometric topology or structured or unstructured mesh.

The retarded potentials are the compression and shear-rotation energies per unit of mass accumulated from a reference instant, here $0$, up to the instant $t^o$:
\begin{eqnarray}
\displaystyle{   \phi^o = - \int_0^{t^o} \: c_l^2 \: \nabla \cdot \mathbf V \:  d\tau;  \:\:\:\:\:\:\:\:\:\:
 \bm \psi^o = - \int_0^{t^o} \:  c_t^2  \: \nabla \times \mathbf V \: d\tau } 
\label{disint}
\end{eqnarray}

At each instant, these retarded potentials are updated with $d \phi$ and $d \bm \psi$. However, these accumulations are complete only in the case where the medium is perfectly elastic. Real waves propagate with an attenuation which depends on the medium; in a Newtonian fluid the transverse waves are attenuated in a elapsed time of the order of $10^{-10}$ to $ 10^{-12}$ second, where they are dissipated in the form of heat. The term $\left(\bm \psi^o - dt \: c_t^2 \: \nabla \times \mathbf V \right)$ must be replaced in this case by $- \nu \nabla \times \mathbf V$, where $\nu$ is the kinematic viscosity. In the general case the discrete equation of motion then takes the form:
\begin{eqnarray}
\left\{
\begin{array}{llllll}
\displaystyle{ \bm \gamma = - \nabla \left( \phi^o - dt \: c_l^2 \:  \nabla \cdot \mathbf V \right) + \nabla \times \left( \bm \psi^o - dt \: c_t^2 \: \nabla \times \mathbf V \right) +  \bm g_s  } \\  \\
\displaystyle{ \left( 1 -\alpha_l \right) \: \phi^o - c_l^2 \: dt \:  \nabla \cdot \mathbf V \longmapsto \phi^o } \\ \\
\displaystyle{\left( 1 -\alpha_t \right) \: \bm \psi^o - c_t^2 \: dt \: \nabla \times \mathbf V \longmapsto \bm \psi^o}
\end{array}
\right.
\label{discrete}
\end{eqnarray}
where $\alpha_l$ and $\alpha_t$ are the attenuation factors of the longitudinal and transverse waves. The source term is written as a Helmholtz-Hodge decomposition, $\mathbf g_s = - \nabla \phi_s + \nabla \times \bm \psi_s$. The $\longmapsto$ symbol represents the upgrade of the potentials from time $t^o$ to time $t$.

It is necessary to underline the autonomous character of the equation (\ref{discrete}); indeed, the mass conservation equation is not associated with the motion equation, as it is for the Navier-Stokes equation. Mass or density does not appear in equation (\ref{discrete}), but both energy and mass are conserved. Finally, it does not contain any constitutive law.

The equivalence between mass and energy can be found through a simple analysis in order of magnitude. If the transverse waves are very quickly attenuated in an isotropic fluid, the longitudinal waves propagate at celerity $c = c_l$, the speed of sound. By considering the quantity $\phi^o - dt \: c_l^2 \: \nabla \cdot \mathbf V$, it is possible to fix the order of magnitude of the second term; if $\mathbf V$ is of order of magnitude of $v$, its divergence is of order $v / d$, and if the wave travels distance $d$ over time period $dt$, this term is then order of magnitude of $c \: v$ and the compression energy per unit mass $\phi^o \propto c \: v$. By writing the energy $e = m \: \phi^o$ with $m$ the mass in motion, we obtain $e = m \: c \: v$ and, if $v = c$, we find the formula of the special theory of relativity $e = m \: c_0^2$ for light. Solving a wave propagation problem using the discrete equation (\ref{discrete}) provides exactly this result.

Consider the special case where $c_l = c_t = c$ and apply the vector calculus formula:
\begin{eqnarray}
\displaystyle{\nabla^2 \mathbf V = \nabla \nabla \cdot \mathbf V - \nabla \times \nabla \times \mathbf V }
\label{gradiv}
\end{eqnarray}

By using the definition of the displacement $ \mathbf U = \mathbf U^o + \mathbf V \: dt$ we obtain a form which contains the retarded potentials at the second member:
\begin{eqnarray}
\displaystyle{  \frac{d^2 \mathbf U}{d t^2} - c^2 \: \nabla^2 \mathbf U = - \nabla \phi^o + \nabla \times \bm \psi^o }
\label{dislumin2}
\end{eqnarray}

The first member of this equation is a d'Alembertian $\dalembert \mathbf U$:
\begin{eqnarray}
\displaystyle{ \frac{1}{c^2} \: \frac{\partial^2 \mathbf U}{\partial t^2} - \nabla^2 \mathbf U  }
\label{dislum}
\end{eqnarray}

Thus the discrete equation of motion is, in fact, an equation of propagation of waves at velocity $c$. Note, however, that $c_l$ and $c_t$ are different in the general case and the relation (\ref{gradiv}) is no longer applicable. The second Lam{\'e} coefficient $\nu$ for solids becomes the kinematic viscosity for fluids, which does not conserve the shear energy.

\textcolor{blue}{\subsubsection{Mass conservation} }

Mass or density is not a variable in discrete mechanics; it is replaced by the total energy per unit mass, $\phi^o$ for compression energy and $\bm \psi^o$ for shear energy. Equation (\ref{discrete}) is autonomous: the variables $(\mathbf V, \phi^o, \bm \psi^o)$ and the celerities are sufficient to describe any problem in fluid or solid mechanics.
The conservation of mass is, however, an essential principle in mechanics; the Lagrangian law of conservation, which depends only on the divergence of velocity:
\begin{eqnarray}
\displaystyle{ \frac{d \rho}{d t} = - \rho \:  \nabla \cdot  \mathbf V  }
\label{massconv}
\end{eqnarray}

It is necessary to dissociate the variations in density due to the changes to the variables of the problem, here velocity, from those related to the advection of the media. This last, very different, phase can be solved in multiple ways: (i) by transforming the particle derivative into a time derivative or (ii) using {\it a posteriori} one of the many Eulerian or Lagrangian methodologies of phase transport (Volume Of Fluid, Moment Of Fluid, Level-Set, Front-Tracking, etc.).

The law (\ref {massconv}) can thus be integrated on a trajectory; the divergence of the local velocity of this law is also that of equation (\ref{discrete}), updated at instant $t^o + dt$; explicit integration provides the solution:
\begin{eqnarray}
\displaystyle{ \rho = \rho^o \:  e^{\displaystyle{- \nabla \cdot  \mathbf V \: dt } } }
\label{masssol}
\end{eqnarray}

Although $\rho$ is not associated with the equation of motion, it can be computed {\it a posteriori} and be the object of a possible transport without diffusion. However, this calculation is not a resolution of the continuity equation but merely an update of the density from $\nabla \cdot \mathbf V$. Both for compressible and incompressible flows, the conservation of mass is ensured intrinsically by equation (\ref{discrete}) through that of the energy per unit of mass.

Pressure $p$ is not a retained quantity either, it is replaced by the scalar potential $\phi$. It can also be calculated in the form $p = \rho \: \phi$ in the case of ideal gases for an isothermal evolution or by $p = \rho^{\gamma} \: \phi $ for an isentropic evolution. In general, the constitutive law $p(\rho, T)$ intervenes at this level simply by assigning to $\phi$ a definition compatible with the behavior of the media considered. The equation of motion remains a generic kinematic law independent of any constitutive relation.

\textcolor{blue}{\subsubsection{Inertia} }

Inertia is a particularly important concept in fluid mechanics because it conditions the chaotic or even turbulent evolutions of certain solutions. The physical analysis set out in reference \cite{Cal20c} shows that it is possible to express inertia starting from a quantity named potential of Bernoulli $\phi_B = \| \mathbf V \|^2/2$ in the form of a Helmhlotz-Hodge decomposition. This scalar potential is defined in all space and inertia appears in discrete mechanics as its curvature; the notion of inertia is still complex in the current view of mechanics.

In the framework of continuum mechanics, the inertial term coming from the material derivative is written $\mathbf V \cdot \nabla \mathbf V$ or $\nabla \cdot \left (\mathbf V \otimes \mathbf V \right) - \mathbf V \: \nabla \cdot \mathbf V$ or $\nabla (\| \mathbf V \|^2/2) - \mathbf V \times \nabla \times \mathbf V$. The last term, the vector of Lamb \cite{Lam93}, \cite{Ham08}, is also not adapted to a discrete description of mechanics \cite{Cal19a} since it is not a dual curl. It is therefore necessary to characterize the inertia in a different way and to rewrite the expression of the material derivative starting from potentials $\phi_i$, the inertial scalar potential and $\bm \psi_i$ the inertial vector potential.
Physically, the variation of velocity over time on the segment $\Gamma$ with respect to the material derivative, i.e. following the segment during its motion, is due to two actions, a direct one associated with the scalar potentials at both extremities and another induced by the circulation of the vector potential along the contour $\Sigma$. The material derivative is fixed by the accelerations imposed from the outside and thus the velocity variation is reduced by a quantity representing inertia. The intrinsic acceleration of the material medium or the particle then takes the form:
\begin{eqnarray}
\displaystyle{ \bm \gamma = \frac{d \mathbf V }{dt}  \equiv \frac{\partial \mathbf V}{\partial t} + \nabla \phi_i - \nabla \times \bm \psi_i = \frac{\partial \mathbf V}{\partial t} + \nabla \left(  \frac{\| \mathbf V \|^2}{2}  \right) -  \nabla \times \left(  \frac{\| \mathbf V \|^2}{2} \: \mathbf n  \right)}
\label{material}
\end{eqnarray}

The application of the divergence and dual curl operators to the two terms of inertia makes it possible to eliminate one of these two terms this is an important difference with the mechanics of continuous media.

\textcolor{blue}{\subsubsection{Invariances and conservation laws} }

Equations in physics are often written as Lagrangians or Hamiltonians. Noether's theorem \cite{Noe11} establishes the equivalences between the Lagrangian invariances of a system and the conservation laws.
The discrete equation can be seen as the sum of two Lagrangians which translate the exchanges between kinetic energy and potential energy like oscillators. The retarded scalar potential $ \phi ^ o $ represents the potential energy and the term $dt \: c_l^2 \; \nabla \cdot \mathbf V $ materializes kinetic energy; in the same way, $\bm \psi^o$ and $dt \: c_t^2 \: \nabla \times \mathbf V$ form the oscillator of the effects of rotation.

The vector equation (\ref{discrete}) is invariant to translational and rotational motions. Galilean invariance results in the fact that a uniform translational motion at constant velocity does not change the equation. This property has been extended in discrete mechanics to rotational motions at constant angular velocity \cite{Cal20c}. The theorem of Noether serves to conclude respectively on the conservation of momentum (here the acceleration) and on that of angular momentum (here per unit of mass). The space is then qualified as homogeneous and isotropic.

The invariance by translation in time expresses the fact that time passes uniformly and that the laws of physics do not depend on it. It reflects the conservation of energy. The discrete equation of motion is a conservation of acceleration but also of mechanical energy. The addition of other Lagrangians on thermal or electromagnetic energy would extend the scope of this equation.

\textcolor{blue}{\subsection{Numerical framework} }

\textcolor{blue}{\subsubsection{Come back to discrete formulation } }

The return to a discrete view of the equation of motion occurs through natural operators ${\rm DIV}$ and ${\rm GRAD}$, initially described by M. Shahkov \cite{Sha96} for transport equations and then extended to the curl \cite{Hym97}. Since then, many works of the mimetic method \cite{Lip14} type applied to different equations (Navier-Stokes, Maxwell, etc.) have shown its efficiency by giving a conservative interpretation of existing methodologies. Mimetic methods are robust and accurate methods which conserve the fundamental properties of equations. At the same time, methods based on differential geometry and external calculus \cite{Des05, Mey03} in a discrete form (DEC, Discrete Exterior Calculus) have been developed over several decades. It is possible to use similar formulations to extract the potentials and the components of a vector on polyhedral meshes by Discrete Helmholtz-Hodge Decomposition \cite{Ahu07, Lem15}. Other variants of mimetic methods have recently been developed, for example for spectral \cite{Pal17} or finite difference \cite{Abb08} methods to solve the Navier-Stokes equation.

The mimetic method is particularly well suited to the discrete equation of motion, written {\it a priori} as a Helmholtz-Hodge decomposition of acceleration. The numerical formulation, albeit applying to a different physical model, is very similar to these methodologies. We will also use operators from the mimetic methodology to transform equation (\ref{discrete}) into a discrete equation where the primary operators are written ${\mathcal GRAD}$, ${\mathcal CURL}$, ${\mathcal DIV}$ for the primal grid and $\widetilde{\mathcal GRAD}$, $\widetilde{\mathcal CURL}$, $\widetilde {\mathcal DIV}$ for the derived dual operators.

The objective of the mimetic method is to create discrete approximations that preserve important properties of continuum equations on general polygonal and polyhedral meshes.
First, we focus on the discretization of differential operators. This phase, called reduction, consists in discretizing the continuous fields into discrete fields. The integral over a polyhedral cell is the sum of the integral over its simplexes. The reduction operator does not introduce any error in the sense that it commutes with respect to differentiation; this result can be obtained from the generalized Stokes theorem.
The second phase, called reconstruction, allows an approximation of the constitutive relations. It allows the approximate representation of dual differential operators. The design of the reconstruction operator defines the convergence rate of the numerical method. Numerical methods exist that use the mimetic framework; we can mention the mimetic finite difference, the finite element method or the spectral element method.

The primal manifold ${\mathcal M}$ is composed of $d_0$ vertices, oriented $d_1$ edges $\Gamma$, $d_2$ facets of normals $\mathbf n$ and $d_3$ cells. The dual mesh $\widetilde{\mathcal M}$ includes $d_4$ dual facets including $d_4 = 2 \: d_1$ in two dimensions and $d_0$ dual volumes.
Using mimetic formalism, the discrete equation of motion becomes:
\begin{eqnarray}
\left\{
\begin{array}{llllll}
\displaystyle{ \bm \gamma = - {\mathcal GRAD} \left( \phi^o - dt \: c_l^2 \:  \widetilde{{\mathcal DIV}} \: \widetilde{\mathbf V} \right) + \widetilde{{\mathcal CURL}} \left( \widetilde{\bm \psi^o} - dt \: c_t^2 \: {\mathcal CURL} \: \mathbf V \right) +  \bm g_s  } \\  \\
\displaystyle{ \left( 1 -\alpha_l \right) \: \phi^o - c_l^2 \: dt \:  \widetilde{{\mathcal DIV}} \: \widetilde{\mathbf V} \longmapsto \phi^o } \\ \\
\displaystyle{\left( 1 -\alpha_t \right) \: \widetilde{\bm \psi^o} - c_t^2 \: dt \: {\mathcal CURL} \: \mathbf V \longmapsto \widetilde{\bm \psi^o}}
\end{array}
\right.
\label{discreted}
\end{eqnarray}

This form causes confusion, especially on the continuous curl operator, $\nabla \times$, as we do not know whether it applies to a vector or to a pseudo-vector. These discrete operators mimic the properties of the continuum \cite{Lip14}, in particular $\widetilde {\mathcal DIV} \: \widetilde{\mathcal CURL} \: \widetilde {\bm \psi} = 0$ and $ {\mathcal CURL} \: {\mathcal GRAD} \: \phi = 0 $. The particle derivative is expressed in the same way, with the same operators of a Helmholtz-Hodge decomposition: 
\begin{eqnarray}
\displaystyle{ \bm \gamma  = \frac{\partial \mathbf V}{\partial t} + {\mathcal GRAD} \left(  \frac{\| \mathbf V \|^2}{2}  \right) -  \widetilde{{\mathcal CURL}} \left(  \widetilde{ \frac{\| \mathbf V \|^2}{2} }  \right)}
\label{materiald}
\end{eqnarray}

The inertial potential \cite {Cal20c}, $\phi_i = \| \mathbf V \|^2/2$, is semi-implied in time for an introduction of the inertial term within the linear system.

The time-stepping procedure is close to that of classical methods: the time derivative $\partial \mathbf V / \partial t \approx \delta \mathbf V / \delta t$ is discretized using a second-order Gear scheme. What is appreciably different, however, is the presence of time lapse $dt$ between two mechanical balances of the system. It must be strictly the same as the time step of the temporal discretization, $\delta t = dt$. This is one of the essential properties of the physical model presented. Discrete equation (\ref{discreted}) makes it possible to understand the physics of phenomena at all time scales and to capture waves at very high frequencies, those of light for example, or to represent stationary flows. The choice of $dt$ by the user must simply be compatible with the physics he or she wishes to simulate.

\textcolor{blue}{\subsubsection{Discrete differential operators} }

The resolution of a linear system whose unknowns are the components $\mathbf V$ of velocity is carried out directly from the discrete system (\ref{discreted}). The four operators of this equation, divergence, gradient, primal and dual curls, are simply formulated implicitly within the linear system. In order to make their description clearer, an explicit version is given. Beforehand, it is essential to note that the physical properties, longitudinal celerity and kinematic viscosity $\nu$ which replaces the grouping $dt \: c_t^2$ for a fluid, are constant on their respective locations, the vertex for $c_l$ and the barycenter of the facet for $\nu$. In this way, quantity $dt \: c_l^2 \: \widetilde {{\mathcal DIV}} \: \widetilde {\mathbf V} $ is constant locally and $ \nu \: {\mathcal CURL} \: \mathbf V $ is constant on each facet.

The gradient of a scalar, $ {\mathcal GRAD} \: \phi$ for example, is a direct operation which can be interpreted as the difference in potential at the vertices, i.e. $ {\mathcal GRAD} \: \phi = ( \phi_b - \phi_a) / d$. This quantity represents the component of the gradient on $\Gamma $, but not the gradient vector in space. It should be remembered that the sum of accelerations is carried out on segment $\Gamma$ and that at no time is it necessary to resort to a representation in terms of a vector. Of course, the operator $\nabla \mathbf V$ of continuum mechanics is not required.

The primal curl at the discrete level is represented by $\bm \psi$, a scalar located on facet $\mathcal S$, obtained from the circulation of the components of velocity on the primal contour $\Gamma^*$; Stokes' theorem specifies that the curl of a vector on surface $\mathcal S$ can be calculated from its components on its contour. Quantity $\bm \psi = \nu \: {\mathcal CURL} \: \mathbf V$ is thus a piecewise constant on the oriented facet. Like the discrete gradient, the primal curl is an exact operation.

The dual curl resulting from the transformation of the primal curl $\bm \psi = \nu \: {\mathcal CURL} \: \mathbf V$ corresponds to the reconstruction phase, which is accompanied by errors due to interpolations. This face, dual-edge transformation, leads to the dual curl $\widetilde{\bm \psi} $ which allows the calculation of the circulation along the dual $\delta$ contour. The discrete result located on the dual face is then transformed into a vector carried by the primal segment $\Gamma$.
It is useful to reiterate that there is no 2D/3D distinction. In two dimensions of space, for a primal planar geometry, the dual vector $\widetilde{\mathcal{\bm \psi}}$ is not in this flat plane; it is orthogonal to it, and for the planar facets in figure (\ref{primdual}), it is directed by unit vector $\mathbf n$.

Divergence $\widetilde{{\mathcal DIV}} \: \widetilde {\mathbf V}$ is calculated from velocity $\mathbf V$ associated with the segment of the primal mesh, which is transformed into dual velocity $\widetilde{ \mathbf V}$ by associating the areas of the dual facet with it. The integral of the discrete divergence on the dual cell divided by its volume makes it possible to assign this divergence to the corresponding vertex.

In vertex-based potential-circulation, these four operators are then assembled two by two to form the two accelerations, ${\mathcal GRAD} (\widetilde{{\mathcal DIV}}) \: \widetilde{\mathbf V})$ for compression and $ \widetilde{{\mathcal CURL}}({\mathcal CURL} \: \mathbf V)$ for rotation.

\textcolor{blue}{\subsubsection{Properties of operators $ {\mathcal GRAD} ( \widetilde{{\mathcal DIV}} \: \widetilde{\mathbf V})$  and $\widetilde{{\mathcal CURL}} ( {\mathcal CURL} \: \mathbf  V)$   } }

In order to check the properties of each of these operators, the choice of a uniform mesh is adopted to separate the sources of intrinsic errors linked to the operators from those generated by the deformation of the meshes.

\begin{figure}[!ht]
\begin{center}
\includegraphics[width=4.9cm]{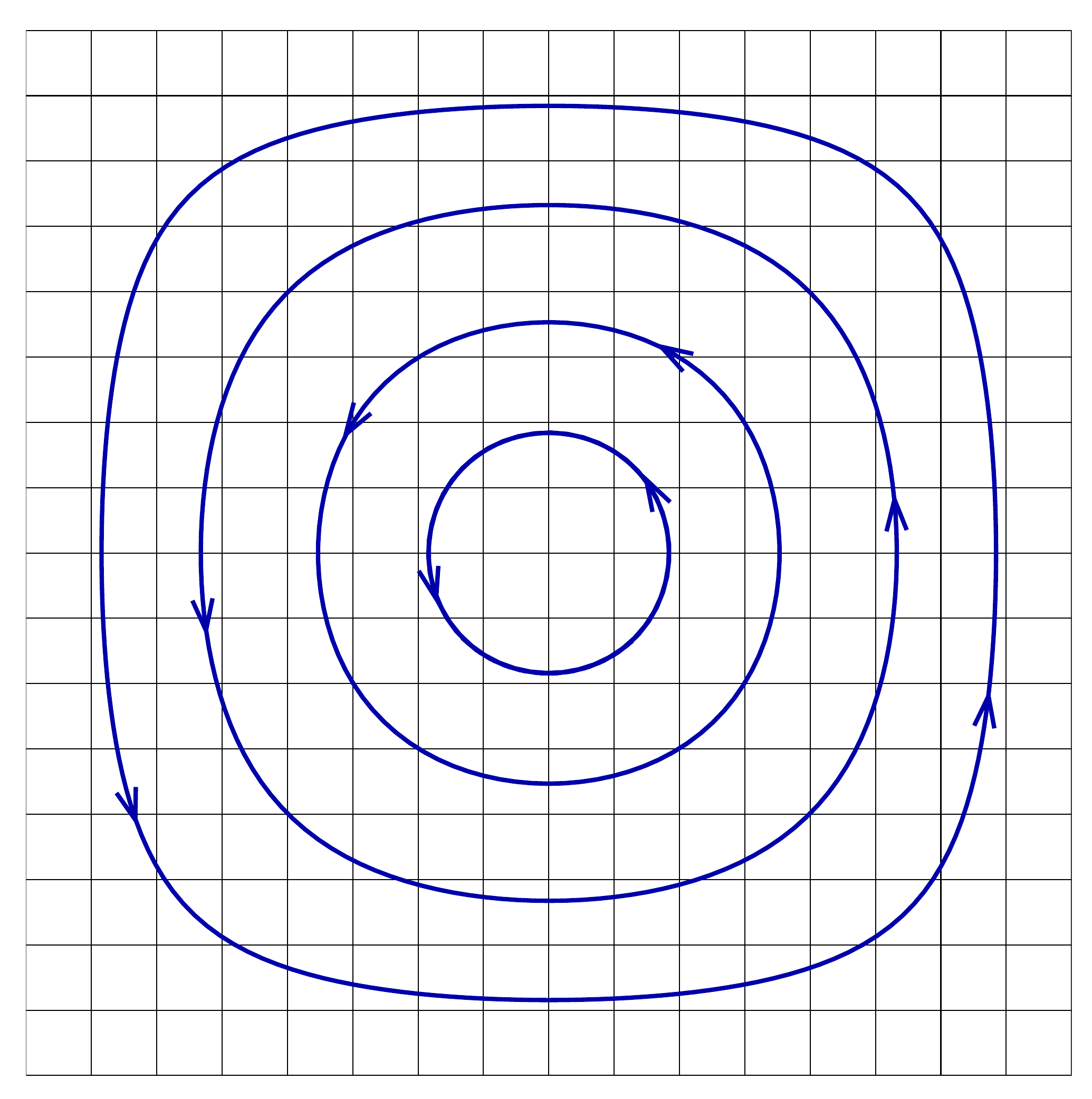}
\includegraphics[width=4.9cm]{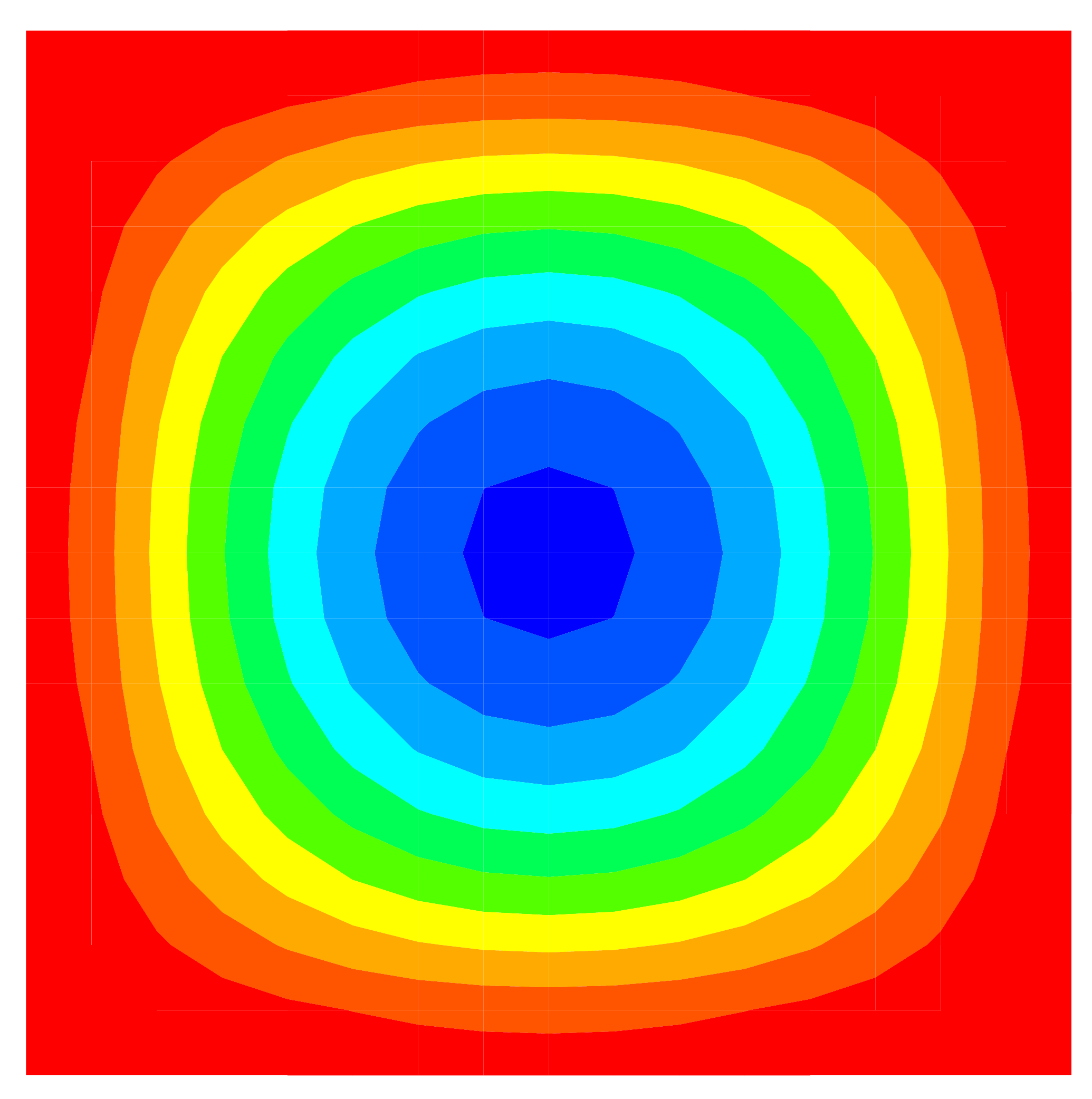}
\includegraphics[width=4.9cm]{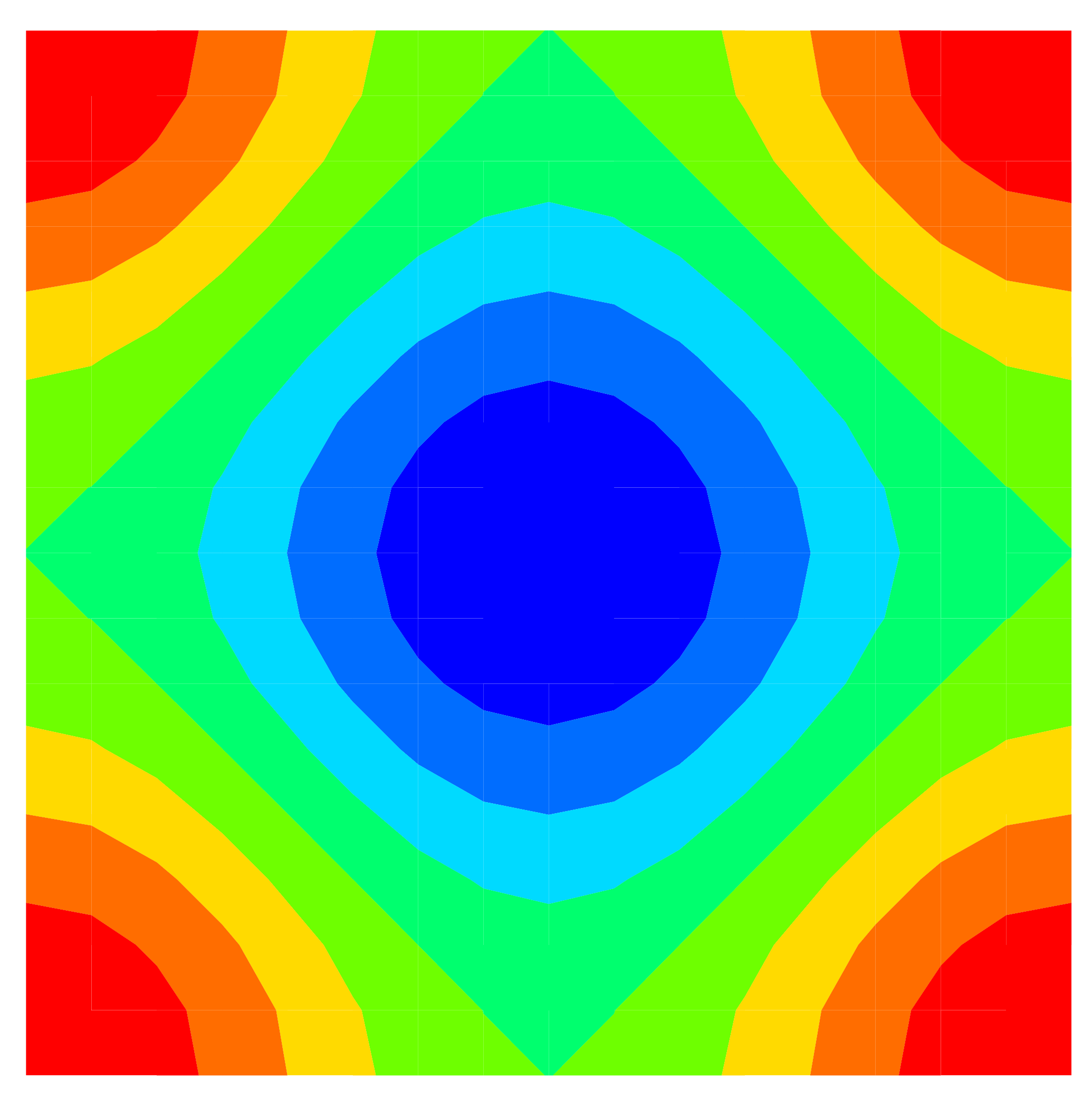}  \\
\hspace{5.mm} (a) \hspace{42.mm} (b) \hspace{42.mm} (c)
\caption{\it Green-Taylor vortex on a Cartesian mesh with 256 cells, 544 edges and 289 vertices, (a) streamlines and mesh, (b) $\bm \psi =  \: \nu \: {\mathcal CURL} \: \mathbf V$, vector potential, (c) $\phi = \phi^o_B - \| \mathbf V \|^2 / 2$, scalar potential. }
\label{green-phi}
\end{center}
\end{figure}

A large number of one-phase and two-phase flow test cases in fluid mechanics show that the convergence rate of solutions is equal to 2.0 in space and time. This is the case for the unsteady Green-Taylor vortex conducted in reference \cite{Cal20a}, but this problem is repeated in numerous publications, \cite{Bel18} for example.
This exact time-dependent solution of the Navier-Stokes equations is used to specify the precision of the operators of the discrete formulation on a domain $\Omega = [-0.5, 0.5]^2$ with time steps small enough to saturate the error in time. Figure (\ref{green-phi}) shows the solution obtained for mesh of $n^2 = 16^2$, (i) streamlines and mesh, (ii) $ \bm \psi $, vector potential and (iii) $ \phi $ scalar potential with $\phi = \phi^o_B - \| \mathbf V \|^2/2$ where $\phi_B$ is Bernoulli potential.

The scalar potential is given at the point of the primal geometric topology and the velocity is reconstructed at the center of the facets only for this graphic representation; the vector potential, also represented on the primal facets, is, on the other hand, always calculated on the barycenter of this representation and is seen like a vector, a component carried by $ \mathbf n $.
The convergence study is carried out for a spatial approximation of $n^2 = 4^2$ to $n^2 = 1024^2$ cells.
\begin{figure}[!ht]
\begin{center}
\includegraphics[width=10.cm]{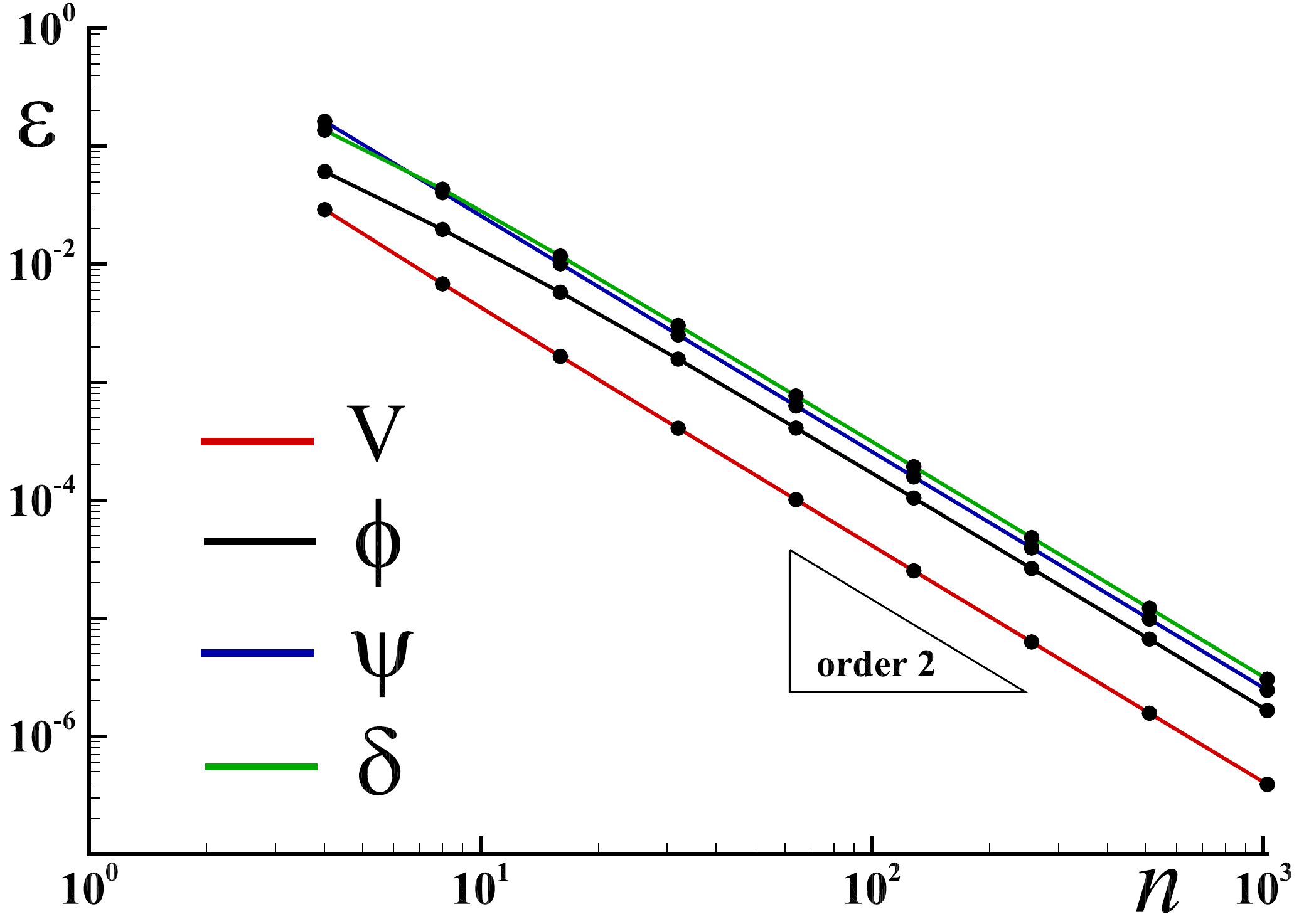} \\
\caption{\it Green-Taylor vortex, spatial convergence  in the error norm $L_2$ for velocity $\mathbf V$,  scalar potential $\phi$, vector potential  $\bm \psi =  \: \nu \: {\mathcal CURL} \: \mathbf V$ and gradient of potential $\bm \delta = {\mathcal GRAD} \phi$. }
\label{green-conv}
\end{center}
\end{figure}

The rate of convergence of the solution on velocity $ \mathbf V $, scalar potential $ \phi $ and vector potential $ \bm \psi $ is equal to $2.0$ (figure \ref{green-conv}). The analysis of the errors of each operator of the discrete equation (\ref{discrete}) is carried out {\it a posteriori} from those on velocity. The scalar potential is upgraded from the divergence of velocity $\phi^o - dt \: c_l^2 \:  \widetilde{{\mathcal DIV}} \: \widetilde{\mathbf V}$  and that of the vector potential deduced from the curl $ \widetilde{\bm \psi^o} = - \nu \: {\mathcal CURL} \: \mathbf V $. For the adopted Cartesian uniform mesh, the errors introduced by these operators do not modify the rate of convergence, also equal to two. Likewise, the computation of the gradient of the scalar potential and of the dual curl of $\bm \psi^o$ saves the spatial convergence to order two.

As it is not possible to conclude on the precise error introduced by each of these operators, the velocity field of the exact solution is projected on the primal mesh, which makes it possible to immediately extract the introduced error. It turns out that (i) the discrete gradient is exact if the scalar potential is itself exact, and that (ii) the primal curl velocity is also exact at machine precision. Thus the second-order errors on the velocity of the numerical simulation are to be attributed to the dual curl and to the velocity divergence.
Of course, the two ${\mathcal GRAD} \: ( \widetilde{{\mathcal DIV}} \: \widetilde{\mathbf V} )$  and $\widetilde{{\mathcal CURL}} ( {\mathcal CURL} \: \mathbf  V)$ are implicit to assemble the linear system, whose only unknown is the velocity component $ \mathbf V $; upgrades on $ \phi $ and $ \bm \psi $ are indeed completely explicit.
\begin{table}[!ht]
\begin{center}
\small
\begin{tabular}{|c|c|c|c|c|}   \hline
    Operator        &       &  &  Property  \\ \hline  \hline
 gradient   &  $\mathcal GRAD \: \phi$     &     $\displaystyle{ \frac{\phi_b - \phi_a}{[\Gamma]} }$   &  exact for any polygonal or polyhedral mesh                 \\ \hline
 primal curl   &  $\mathcal CURL \: \mathbf V$     &     $\displaystyle{ \frac{1}{[\mathcal S]} \sum_{k \in \Gamma^*} \mathbf V_k \: d_k  }$   &  exact for any polygonal or polyhedral mesh                 \\ \hline
 dual curl   &  $\widetilde{{\mathcal CURL}} \: \widetilde{\bm \psi}$     &     $\displaystyle{ \frac{1}{[\Delta]} \sum_{k \in \delta} \widetilde{\bm \psi_k} \: \widetilde{d_k }}$    &  $\mathcal O(h^2)$ for uniform mesh  \\ \hline
 divergence   &  $\widetilde{{\mathcal DIV}} \: \widetilde{\mathbf V}$     &     $\displaystyle{ \frac{1}{[\Omega]} \sum_{k \in \Delta} \widetilde{\mathbf V_k} \: \widetilde{\mathcal{S}_k} }$    &  $\mathcal O(h^2)$ for uniform mesh  \\ \hline
 mimetic   &      &     $\widetilde{\mathcal DIV} \: \widetilde{\mathcal CURL} \: \widetilde{\bm \psi} = 0$    & exactly for any polygonal or polyhedral mesh  \\ \hline
 mimetic   &      &     ${\mathcal CURL} \: {\mathcal GRAD} \: \phi = 0 $    &  exactly for any polygonal or polyhedral mesh  \\ \hline
 orthogonality   &  local     &     ${\mathcal GRAD} \: \phi \cdot \widetilde{{\mathcal CURL}} \: \widetilde{\bm \psi} = 0$    &  exactly for any polygonal or polyhedral mesh  \\ \hline
 $\mathbf V$ polynomial   &  degree $\le 2$   & ${\mathcal GRAD} \: ( \widetilde{{\mathcal DIV}} \: \widetilde{\mathbf V} )$    &  exact for uniform mesh  \\ \hline
  $\mathbf V$ polynomial   &  degree $\le 2$   & $\widetilde{{\mathcal CURL}} ( {\mathcal CURL} \: \mathbf  V)$    &  exact for uniform mesh  \\ \hline
\end{tabular}
\normalsize
\caption{Properties of discrete operators }
\label{properties}
\end{center}
\end{table}

Table (\ref{properties}) presents the main properties of the operators of discrete mechanics. It is of course possible to apply the numerical methodology presented to any geometries tessellated by very deformed meshes and, in this case, the rate of convergence of $ 2.0 $ will not be maintained. The precision of the method according to the quality of the mesh is not the aim of the article. The objective is to know the errors introduced by the processing of the jump conditions for two-phase flows, which it is essential to separate from the errors of the method and those introduced by different considerations linked to the transport of the interfaces, calculation of the curvatures, calculation of the velocity of interfaces, etc.

Finally, we show \cite{Cal20c} that the global structure obtained has important properties \cite{Cal20a}: (i) the vectors ${\mathcal GRAD} \: \phi$ and $\widetilde{{\mathcal CURL}} \: \widetilde{\bm \psi}$ are locally orthogonal, (ii ) the discrete operators mimic the properties of the continuous, $\widetilde{\mathcal DIV} \: \widetilde{\mathcal CURL} \: \widetilde{\bm \psi} = 0$ and ${\mathcal CURL} \: {\mathcal GRAD} \: \phi = 0$ regardless of the regular functions $\phi$ and $\widetilde{\bm \psi}$.

\textcolor{blue}{\subsubsection{Why uniform meshes?} }

The numerical resolution of equations of fluid mechanics for complex physical domains or for two-phase flows presents difficulties of conformity of the mesh with that of the interfaces. But these pitfalls are also observable in problems of mechanics of one-phase fluids. The tessellation of simple surfaces or volumes, a circle for example, is not possible with elements whose faces are based on equilateral triangles or rectangles; the irregularity of the mesh of the physical domain inevitably leads to errors, generally due to the non-planarity of the primal and dual surfaces where the fluxes are calculated.

To obtain this property it is necessary that, for each facet of the primal and dual geometric topologies:
\begin{itemize}[label= \textcolor{blue}{\textbullet}]
\item the facets $\mathcal S$ of unit normals $\mathbf n$ are planar;
\item the segments joining the barycenter of facet $\mathcal S$ in the middle of each segment $\Gamma$ are orthogonal to it, which results in the dual surfaces $\Delta$  also being flat.
\end{itemize}
where $\mathcal S$ and $\Gamma$ are defined in figure (\ref{primdual}).

These conditions can only be met for regular polygonal or polyhedral meshes, such as tessellations based on equilateral triangles or rectangles in two dimensions of space, or regular tetrahedra or polyhedra based on Cartesian hexahedra in three dimensions of space. This class of meshes, qualified here as uniform, is the most capable of producing excellent-quality results in many fluid mechanics problems.

In order to overcome the problem of the conformity of the interfaces between different environments or geometric domains which are sometimes difficult or even impossible to tessellate with these uniform meshes, the proposed method consists in imposing conditions on the cells cut by an interface or a surface of the domain. There are many similar techniques designated by the term ``Immersed Boundary Methods'' \cite{Mit05}. Other methods, corresponding to the penalization of some of the terms of the equation of motion, also make it possible to obtain orders of convergence of one or two \cite{Arq84, Ang99, Ang12,  Cal15c, Tav20}. The jump conditions proposed here are specific to the discrete formulation of the equation of motion, whose physical foundations are outlined elsewhere \cite{Cal19a, Cal20a}.

The numerical methodology associated with discrete mechanics applies to non-regular polygonal or polyhedral meshes with any number of faces which do not have the property of flatness. In this case, the numerical solution is not free from errors, even for solutions which correspond to polynomials of degree less than or equal to two; in general the convergence of the solution is of order two in space.

\textcolor{blue}{\section{Treatment of jump conditions} }

The treatment of the conditions of jump in discrete mechanics differs appreciably from that adopted for the continuous mediums. Several procedures explain these differences: (i) mass or density is removed from the equation of motion except when Archimedean acceleration is activated in an adequate potential, (ii) the effects of compression and rotation are shown by orthogonal terms, and (iii) the jump terms conform to the derivation of all the terms of the equation, a formal Helmholtz-Hodge decomposition.

The standard treatment of two-phase flows generally shows spurious currents, especially when capillary effects are taken into account through a source term. Reviews devoted to these effects, such as that of Popinet \cite{Pop18}, help to understand these effects due to the modeling of one or other of the physical phenomena included. More precise analyses on the restrictions of the chosen time step are also available in the literature, such as that of Denner \cite{Den15b} on surface currents that are very sensitive to time and space steps. All these problems would require specific analysis within the framework of discrete mechanics. For example, if the capillary source term is not a true gradient compensated by the pressure gradient, it is not possible to obtain a stationary solution without parasitic currents. Unfortunately, the defects of the curvature calculation generate rotational.
\begin{figure}[!ht]
\begin{center}
\includegraphics[width=5.cm]{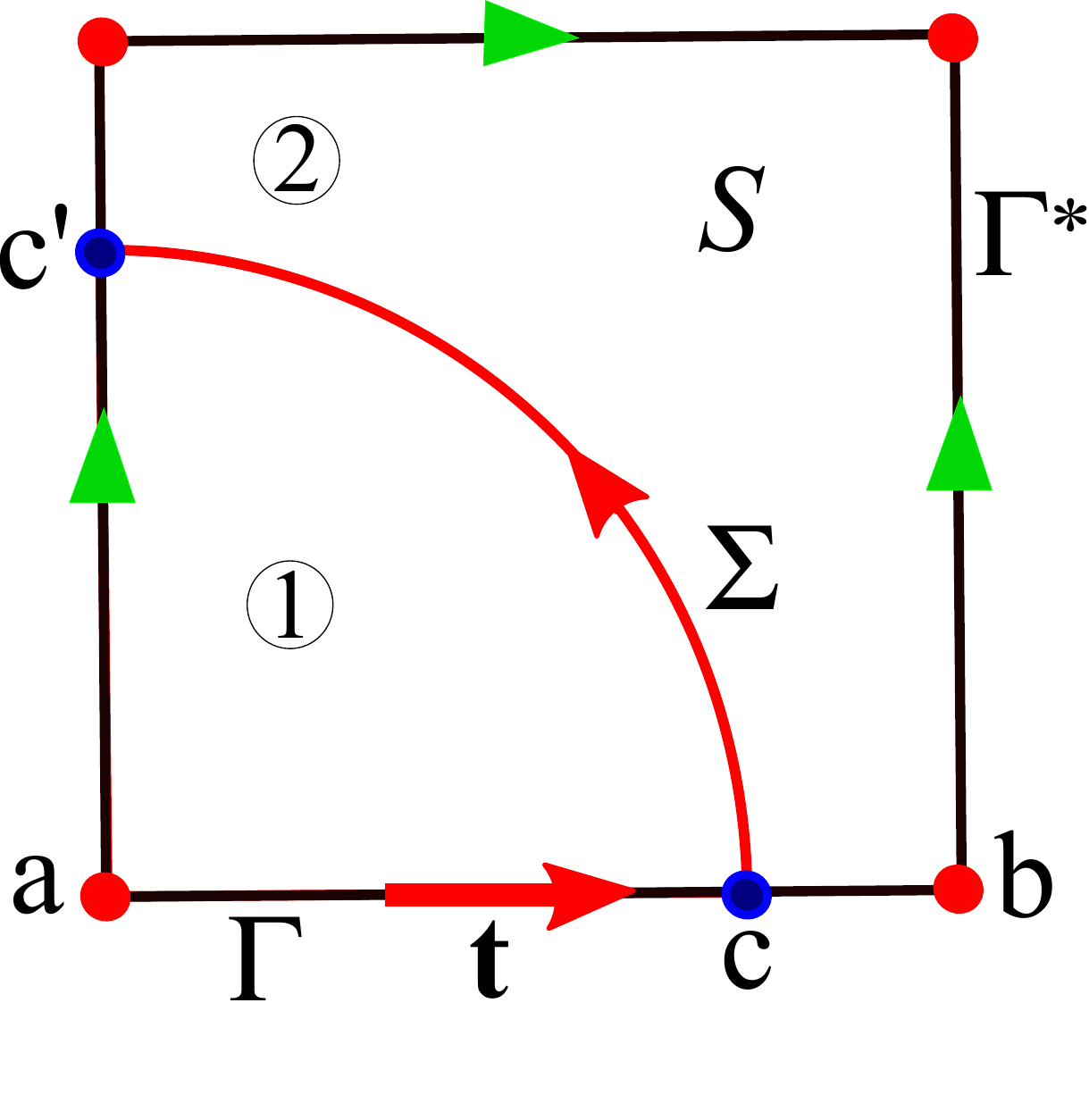}
\vspace{-5.mm}
\includegraphics[width=4.5cm]{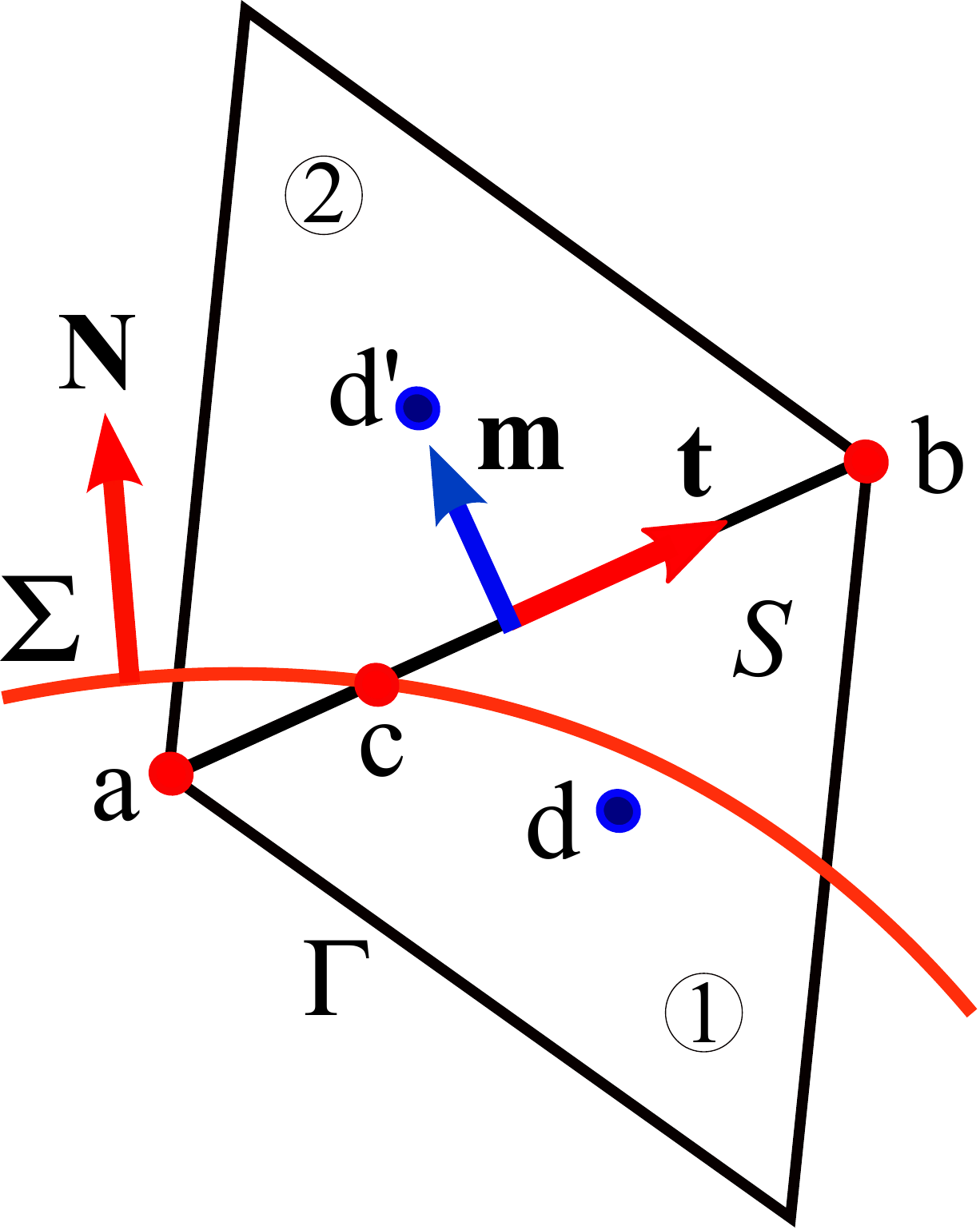} \\
\hspace{0.mm} (a) \hspace{40.mm} (b)  
\caption{\it Discrete geometric topology:  (a) the surface $\mathcal S$ can be composed of two domains {\rm \textcircled{\scriptsize 1}} and {\rm \textcircled{\scriptsize 2}} such that $\mathcal S = \mathcal S_1 \cup \mathcal S_2$. The curvilinear oriented segment $\Sigma$ materializes the trace on $\mathcal S$ of the interface between the two domains cutting the edges in $c$ and $c'$. (b) the interface $\Sigma$ of normal $\mathbf N$ passes through two coplanar facets and the $\Gamma$ segment at the point $c$; $d$ and $d'$ are the centroids of cells and $\mathbf n \times \mathbf t$ the unit vector of the direct local coordinate system $(\mathbf m, \mathbf n, \mathbf t)$.}
\label{stencils}
\end{center}
\end{figure}

The treatment of density jumps $[[\rho]]$, viscosity $[[\nu]]$ or capillary pressure $[[p_c]]$ is not performed in the same way in discrete mechanics. The first of these is assigned to the $\Gamma$ segment of the primal geometric topology, the viscosity jump is associated with the centroids of the cells and the capillary pressure jump is implemented by the two terms of a Hodge-Helmholtz decomposition of the capillary potential.

\paragraph{\textcolor{blue}{Density jump:} }

The density is associated with the scalar potential and the pressure by the relation $\phi^o = p / \rho$. Like these last two quantities the density is defined on each vertex of the primal geometry. If the density is constant and uniform in the whole domain, the pressure will be equal to the potential $\phi^o$ up to a multiplicative constant. For a two-phase flow where the fluids have constant densities $\rho_1$ and $\rho_2$, only the advection of the phases will be taken into account. In the case where the density varies for other reasons, temperature or pressure variations for example, it will have to be upgraded from the conservation of the mass:
\begin{eqnarray}
\displaystyle{  \frac{d \rho}{d t} = - \rho \: \nabla \cdot \mathbf V  }
\label{masvol}
\end{eqnarray}

For the only two-phase flow cases considered here the density jump $[[\rho]]$ will naturally be equal to $(\rho_b - \rho_a)$ for all the segments cut by the interface $\Sigma$.

\paragraph{\textcolor{blue}{Viscosity jump:} }

The formulation (\ref{discrete}) itself sets the condition at the boundary between two facets whose kinematic viscosities $\nu_1$ and $\nu_2$ are different and constant over them. If the velocity fields are respectively $\mathbf V_1$ and $\mathbf V_2$, the condition at the interface is given by:
\begin{eqnarray}
\displaystyle{   \nu_1 \: \mathcal CURL \: \mathbf V_1 = \nu_2 \:  \mathcal CURL \: \mathbf V_2 }
\label{condlim}
\end{eqnarray}

By definition, the curl of the velocity is constant on each facet and is the same for the viscosities of the product $\nu \: \mathcal CURL \: \mathbf V$ which is none other than the vector potential $\widetilde{\bm \psi}$. The condition (\ref{condlim}) is actually respected locally and implicitly in all points of the domain.

When a facet shown in figure (\ref{stencils}a) is cut by the interface $\Sigma$, it is necessary to compute the value of the mean viscosity $\nu_m$ using the properties of the theorem of Stokes by calculating the velocity of circulation, and so of its components, along contours $\Gamma_1$ and $ \Gamma_2$
\begin{eqnarray}
\displaystyle{  \nu_m \: \mathcal CURL \mathbf V = \frac{1}{[\mathcal S]} \int_{\Gamma_1} \nu_1 \: \mathbf V \cdot \mathbf t \: dl + \frac{1}{[\mathcal S]} \int_{\Gamma_2} \nu_2 \: \mathbf V \cdot \mathbf t \: dl =  \nu_1 \: \mathcal CURL \: \mathbf V_1 +  \nu_2  \: \mathcal CURL \: \mathbf V_2 }
\label{visco}
\end{eqnarray}
where $\Gamma_1$ and $\Gamma_2$ represent the contours of surfaces $\mathcal S_1$ and $\mathcal S_2$ and $ \mathbf V_1$ and $\mathbf V_2$ the tangential components of the velocity on these contours.
Quantity $\zeta_m = [\mathcal S_1] / [\mathcal S]$ will define the partition factor of the phases on the facet $\mathcal S$. The velocity flow along the interface $\Sigma$ will be computed from the velocity on the centroid reconstructed from the components $\mathbf V$, i.e. $(\mathbf V \cdot \mathbf t_{\Sigma})$.
As the components of the velocity on each edge, the viscosities $\nu_1$ and $\nu_2$ on each portion of the facet $\mathcal S$ and the trace of the interface $\Sigma$ inside a cell are known, it is possible to compute the average viscosity $\nu_m$ from the expression (\ref{visco}). It is also possible to obtain an implicit form of $\nu_m \: \mathcal CURL \: \mathbf V$ and integrate it into the expression (\ref{condlim}) to express the operator $\widetilde{\mathcal CURL} (\nu \: \mathcal CURL \: \mathbf V)$ within the algebraic linear system.

\paragraph{\textcolor{blue}{Capillary pressure jump:} }

The jump condition located on edge $\Gamma$ corresponding to the capillary effects is modeled in 3D by the two source terms of a Hodge-Helmholtz decomposition of the capillary acceleration $\bm \gamma_c$:
\begin{eqnarray}
\displaystyle{  \bm \gamma_c =  {\mathcal GRAD}  \left(  \sigma \: \kappa \: \xi \right)  - \widetilde{{\mathcal CURL}}  \left( \widetilde{\sigma \: \kappa \:  \zeta } \right) } 
\label{capil}
\end{eqnarray}
where $\phi_c = \sigma \: \kappa = p_c / \rho$ is the capillary potential  with $\sigma = \gamma / \rho$ the surface tension per unit of mass, $\kappa$ the curvature and  where $\xi$ et $\zeta$ are phase functions located respectively on the points and on the barycenters of the facets.
When the source term is constant, for a static equilibrium for example, it can be interpreted as well as the gradient of a scalar function or as the dual curl of a vector potential. 

Like the scalar potential of the acceleration, the capillary potential is expressed in $m^2 \: s^{-2}$. 
The capillary potential difference between the two fluids $\Delta \phi_c$ is independent of the density, but this is not the case for the capillary pressure which must be computed by  $\Delta p_c =  \rho \: \Delta \phi_c$. 
The capillary pressure jump $p_c$ is defined from quantities located at the points of the primal geometric topology, the potential $\phi_c$ and the density $\rho$. This model belongs to the class of sharp interfaces. 
To carry out a simulation of a two-phase flow it is advisable to redefine the scalar potential $\phi^o = p / \rho$ at each iteration in time in order to take account of the advection of the phases.

Consider the 2D-case of a circle of curvature $\kappa = 1 / R$ placed in the center of a square with side $4 \: R$; the constant surface tension is equal to $\sigma$. When the discrete equation of motion is restricted to only capillary effects it is written:
\begin{eqnarray}
\displaystyle{  \bm \gamma = - {\mathcal GRAD}  \left( \phi^o - dt \: c_l^2 \: \widetilde{\mathcal DIV} \widetilde{ \mathbf V } \right) + {\mathcal GRAD}  \left( \sigma \: \kappa \: \xi \right) }
\label{drop}
\end{eqnarray}

This equation is solved from zero values of the velocity field and the scalar potential. The solution is obtained in a single step with a precision of $\approx 10^{-16}$ compared to the theoretical solution $\phi^o_c = \sigma \: \kappa$.
At the end of the resolution the velocity is zero and of course $\widetilde{\mathcal DIV} \widetilde{ \mathbf V } = 0$.

\paragraph{\textcolor{blue}{Implementation:} }

The general numerical methodology associated with solving two-phase problems whose interfaces $\Sigma$ do not conform to the structured or unstructured meshes used is described below:

\begin{itemize}[label = \textcolor{blue}{\textbullet}]
\item locate the points of the primal mesh that are inside the object (phase $1$) by a ray-tracing method;
\item detect all intersections of the trace of the interface with $\Gamma$ segments and find the intersection points for each facet exactly $\mathcal S$;
\item define the polygons with 3, 4, 5 sides (or more depending on the tessellation) and calculate their areas; check whether the total area of the object is found exactly;
\item compute the occupancy rate of the {\rm \textcircled {\scriptsize 1}} and {\rm \textcircled {\scriptsize 2}} phases on each facet;
\item compute the circulation on the contour of each cell portion crossing the interface $\Sigma$;
\item compute the values of the properties on each intersected element, the density $\rho$ on the vertices, the viscosity $\nu_m$ on the facet and the mean curvature on the segment.
\end{itemize}

All these operations are performed with very efficient differential geometry routines. They will be repeated at each time step for a real simulation where the interface is moving within a fixed mesh. In the case of a Lagrangian approach (ALE for example), the interface is in conformity with the mesh and the jump conditions are not necessary, but the formulation applies nevertheless.

\textcolor{blue}{\section{Numerical results} }

Standard test cases relating to one-phase or two-phase flows have already been carried out \cite{Ang12}, with a similar formulation to that presented, in particular, on the benchmark of a bubble rising in a liquid \cite{Hys09}. Others have been carried out more recently for two-phase flows or fluid-structure interactions \cite{Cal19a, Cal20a}. The results are in agreement with those of the literature.

The numerical solutions shown here generally have a solution of degree less than or equal to two on the velocity which corresponds to $\phi$ and $\bm \psi$ potentials linear or piecewise linear and thus to accelerations ${\mathcal GRAD} \: \phi$ and $\widetilde{{\mathcal CURL}} \: \widetilde{ \bm \psi}$ constant. When each dual $\Delta$ facet is orthogonal to the $\Gamma$ segment, the numerical solution must be exact.

\textcolor{blue}{\subsection{Equilibrium of two phases under gravity} }

\begin{figure}[!ht]
\begin{center}
\includegraphics[width=7.cm]{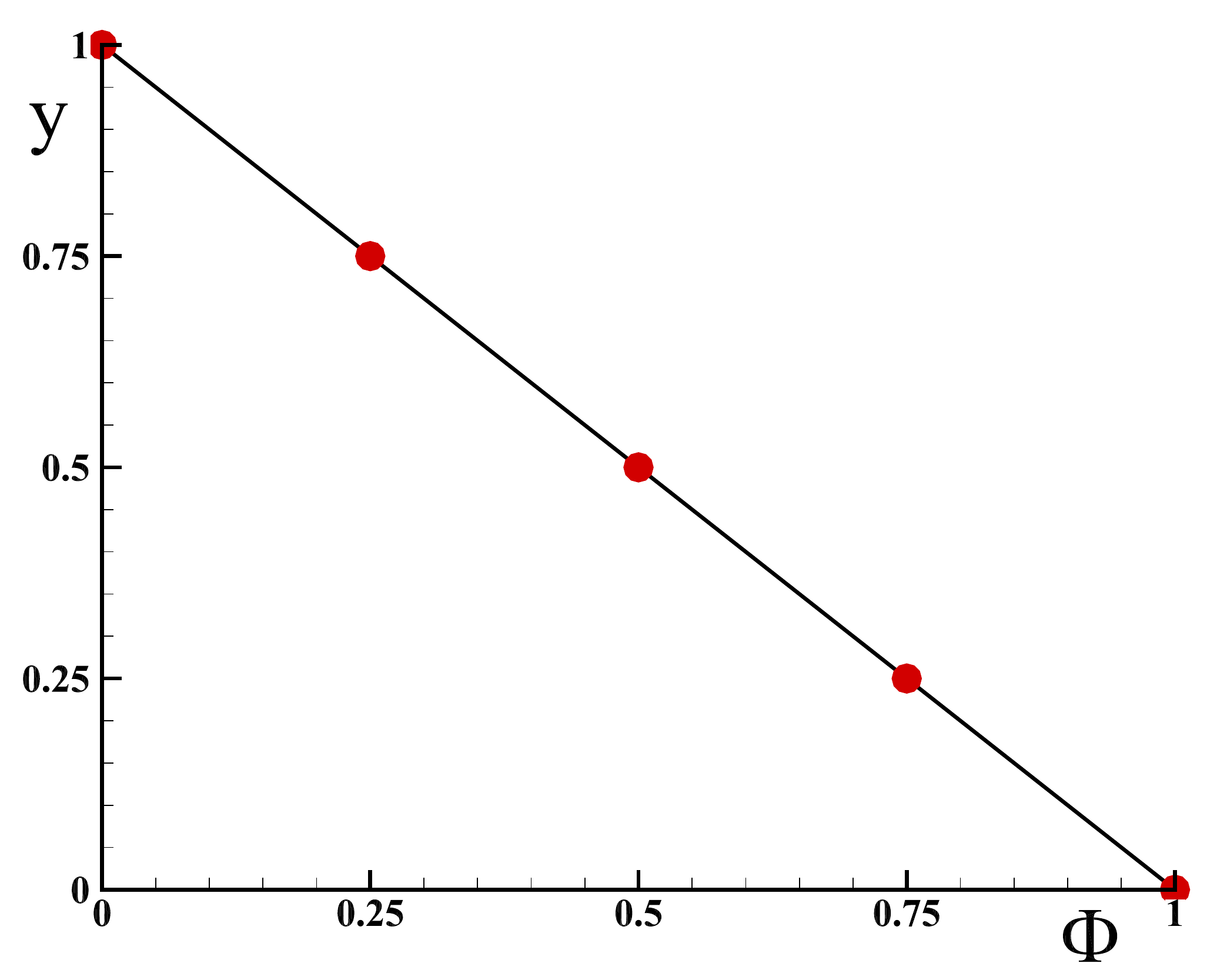}
\includegraphics[width=7.cm]{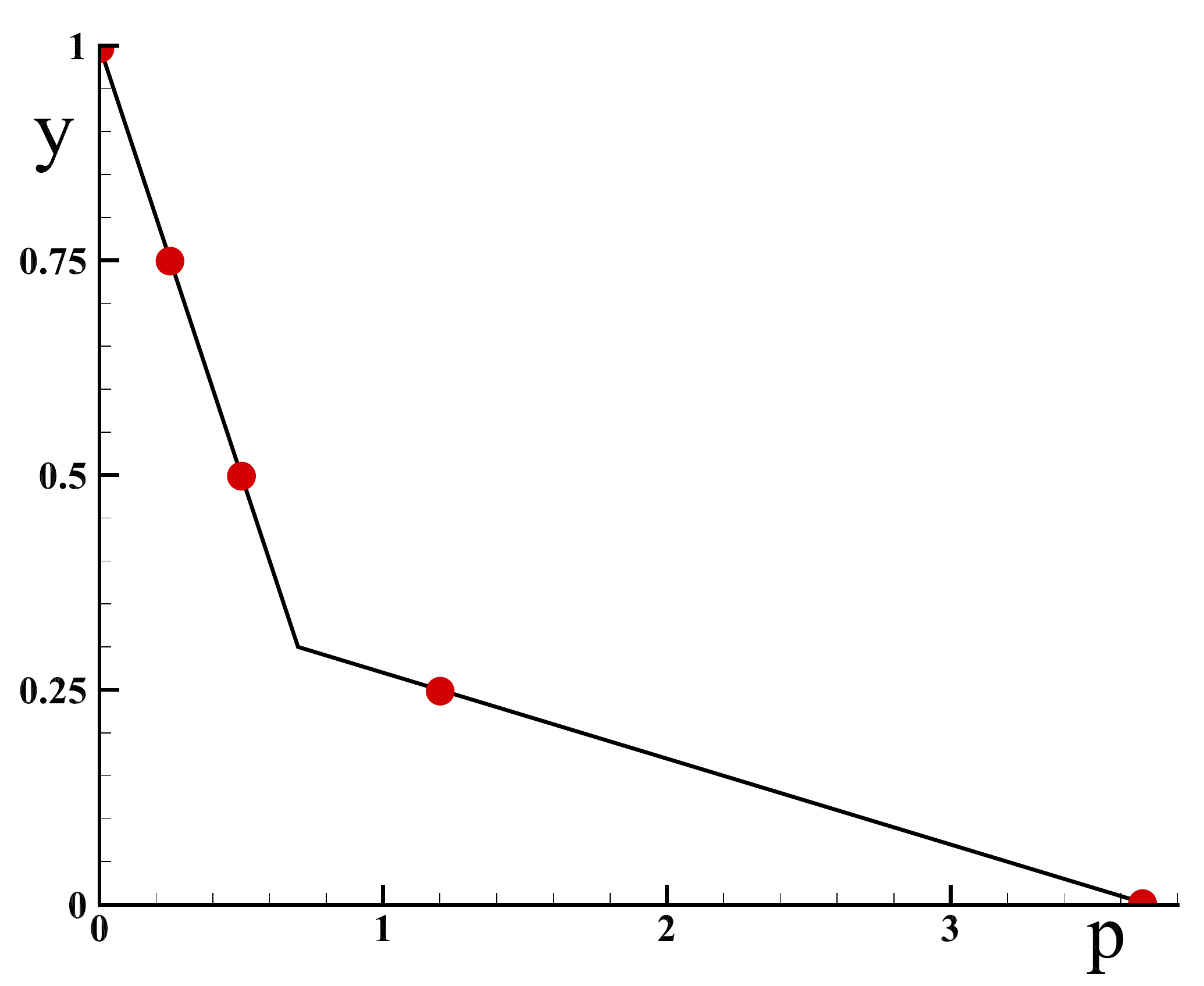}
\caption{\it Equilibrium of two fluids subjected to the influence of gravity; on the left, the evolution of the potential $\phi^o(y)$ and on the right the evolution of the pressure $p(y)$. The theoretical solution is represented by lines and the numerical solution by points; the solution is accurate to the machine precision. }
\label{gravity}
\end{center}
\end{figure}

Consider two fluids of density $\rho_1$ and $\rho_2$ and kinematic viscosity $\nu_1$ and $\nu_2$; the two fluids are superimposed in a cavity of height $H = h_1 + h_2$ where $h_1$ and $h_2$ are the heights occupied by each of the fluids. At the initial moment, the set is subjected to gravity according to $y$, $\mathbf g = -g \: \mathbf k = - g \: \mathbf e_y$ where vector $\mathbf k$ is vertical ascending. The heavier fluid can logically be below, but even if the equilibrium is potentially unstable, the theoretical and numerical solutions are perfectly stable. Initially, the pressure and velocity are zero.
Figure (\ref{gravity}) shows the scalar potential $\phi^o(y)$ solved and the pressure $p(y)$ which is extracted after the explicit upgrade defined by the continuous expression (\ref{upgrade}):
\begin{eqnarray}
\displaystyle{   p_b = p_a - \int_a^c \rho_1 \: \nabla \phi \cdot \mathbf t \: dl  - \int_c^b \rho_2 \: \nabla \phi \cdot \mathbf t \: dl  }
\label{upgrade}
\end{eqnarray}

The system equation (\ref{discrete}) is resolved with a time step of $dt = 10^{12} \: s$ in order to obtain the steady solution in a single step. During the resolution of the linear system, the velocity is directed downwards for the two fluids, but as the incompressibility is applied very strongly, it is null at the end of the algebraic resolution and the solution of problem $ (\phi^o, \mathbf V)$ at equilibrium is obtained instantly with $\mathbf V = 0$; the uniform mesh is composed of $4$ cells. The heights of the two fluids are respectively $h_1 = 0.3 \: m$ and $h_2 = 0.7 \: m$; the heavy fluid placed in the lower part of the cavity has a density equal to $\rho_1 = 10$ and the light fluid a density of $\rho_2 = 1$.

The velocity obtained is at zero divergence $(\widetilde{\mathcal DIV} \widetilde{ \mathbf V } \approx 10^{-16})$ and the pressure corresponds to the theoretical solution sought. Viscosities $\nu_1$ and $\nu_2$ are taken into account only during resolution, as the stationary solution is independent of this physical parameter. The density ratio is chosen as being equal to $\rho_1 / \rho_2 = 10$ and may be arbitrary.

\begin{figure}[!ht]
\begin{center}
\includegraphics[width=6.cm]{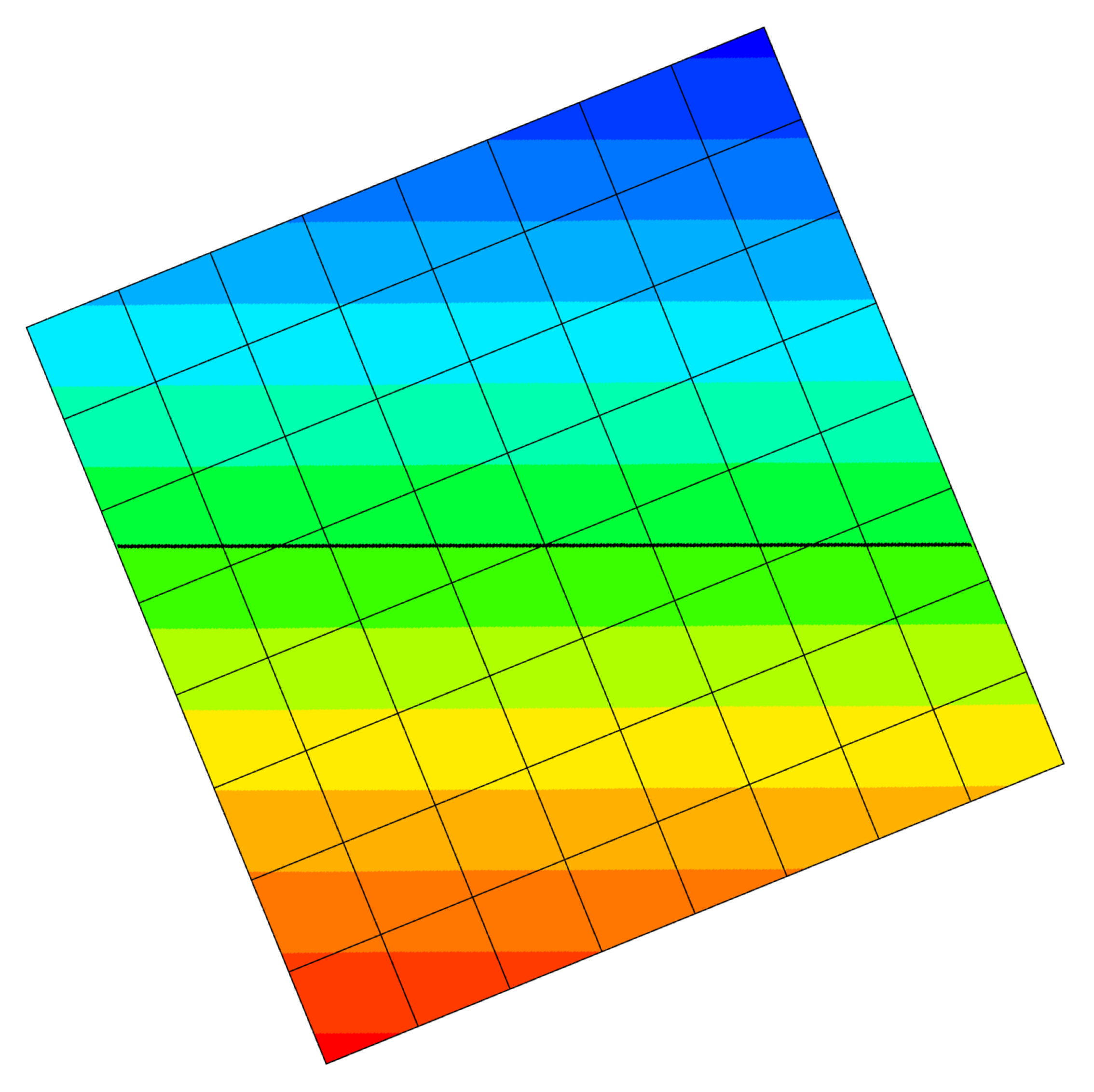}
\includegraphics[width=6.cm]{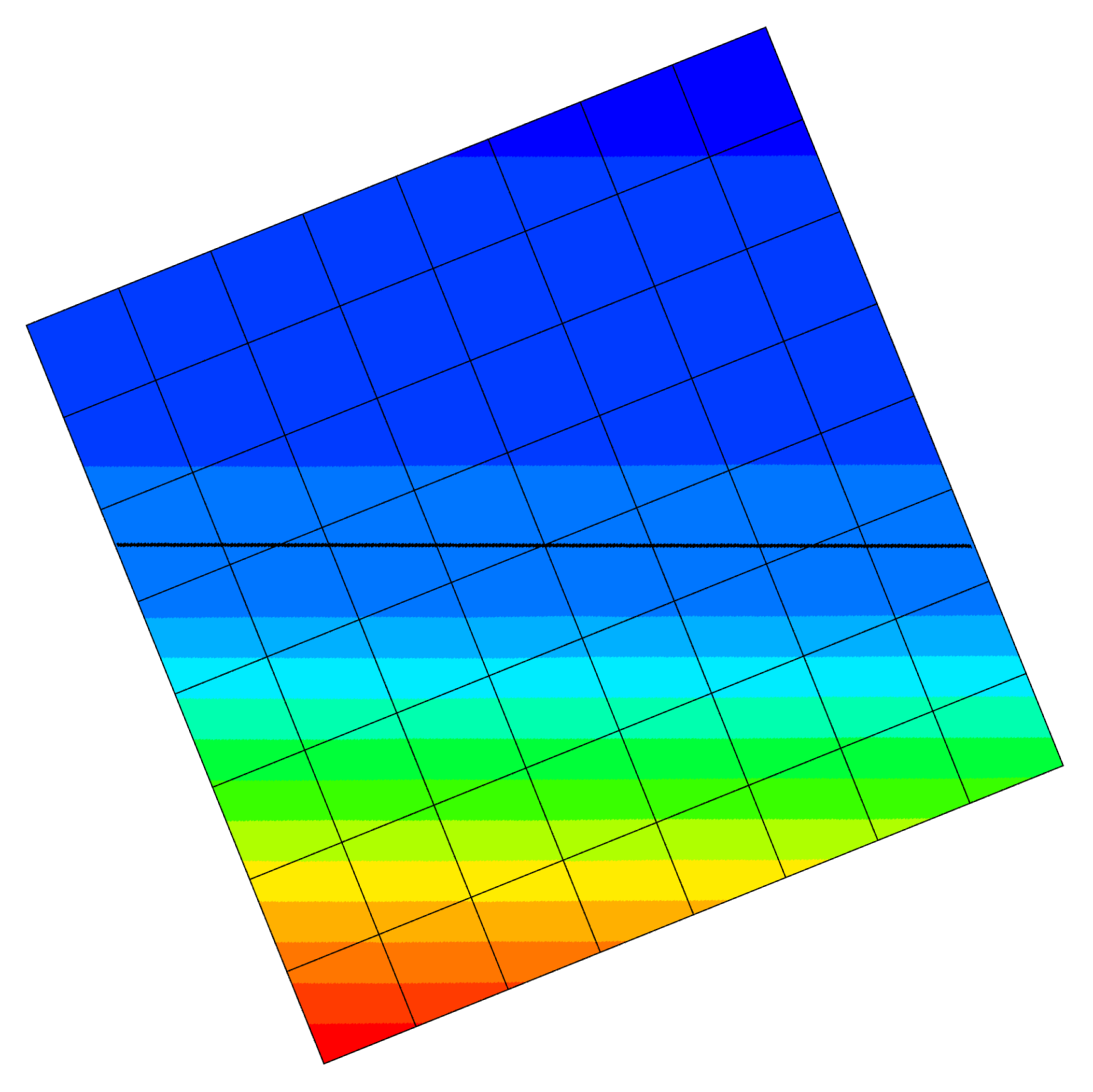}
\caption{\it Equilibrium of two fluids subjected to the influence of gravity; on the left the scalar potential $\phi \in [\pm 0.065] $ and on the right the pressure $ p \in [0, 0715] $; the numerical solution is equal to the theoretical solution to machine precision and the velocity is zero; the isovalues are not drawn in the vicinity of the interface because the graphic interpolation between two points is linear and not, linear by pieces.    }  
\label{gravitation}
\end{center}
\end{figure}

This test case has already been carried out recently using a finite element method \cite{Ili19}, where the analytical solution is found exactly.
One of the advantages of the incompressible formulation presented here  lies in the absence of density in the equation of motion, which makes it possible to obtain a solution on the potential $\phi = p / \rho$ regardless of this.
The same test case, whose mesh does not conform with the interface, is carried out in a square $ [- 0.05 -0.05] ^ 2 $ inclined with a slope of $tg (\theta) = 0.4$, that is, an angle $\theta = 0.3805 \: rd$ and a descending vertical gravity $ \mathbf g = -10 \: \mathbf e_y $. The cavity is filled half with a fluid of density $ \rho_1 = 10 $ and half of density $ \rho_2 = 1 $, the interface is horizontal and the pressure is zero initially. The flow is supposed to be incompressible and the divergence is zero throughout the calculation. The solution can be obtained in an iteration in time or a time step $ dt $; the stationary solution is the same, it is determined with machine precision whatever the adopted uniform mesh. Figure (\ref{gravitation}) shows the numerical solution obtained for a Cartesian mesh of $8^2$ cells.

This case does not present parasitic currents at any time; the interface remains flat during the simulation. When the surface is initially inclined with respect to the horizontal direction, the Archimedes effects generate a movement which attenuates if the viscosity is present to give the previous solution again. This dynamic case of sloshing will not be treated within the framework of this analysis devoted to the treatment of the conditions of jumps, but the solution obtained with the same model can be found elsewhere \cite{Cal19a}, where the oscillation frequencies are compared to the results of linear theory.
It should be noted that the density is not required for this problem, because the equation of the statics of fluids in a continuous medium is written $ - \nabla p + \rho \: \mathbf g = 0 $, while the discrete equation is written $ - {\mathcal GRAD} \: \phi + \mathbf g = 0 $. The quantity $\phi$ is the potential of acceleration, which is not the case of pressure.

In fact, the analytic treatment of this problem does not require any resolution, it is sufficient to note that the vector $\mathbf g$ is constant and that it can derive from a vector potential $\bm \psi_g$ or from a scalar potential $\phi_g$ such as $\mathbf g = {\mathcal GRAD} \: \phi_g$. We can  explicitly calculate the potential on each point of the structured or unstructured mesh:
\begin{eqnarray}
\displaystyle{   \phi_b = \phi_a - \int_a^b  \mathbf g \cdot \mathbf t \: dl    }
\label{upgradegrav}
\end{eqnarray}

It is sufficient to choose a first point whose value will be arbitrary and to traverse the whole of the primal mesh point after point to calculate the potential to a constant and the pressure $p = \rho \: \phi$.

\textcolor{blue}{\subsection{Poiseuille flow in non conform structured mesh} }

The flow of Poiseuille in a planar channel can be simulated using a Cartesian mesh conforming to the geometry of the channel. In this case, the numerical solution is accurate to machine precision.
The test case presented here corresponds to the same planar channel but is inclined in some way in a regular Cartesian mesh. The height channel $h = [\pm 0.1342] $ is bounded by two solid planes inclined at $\theta = 26.56$ from the horizontal. The fluid has a density equal to $\rho$ and a kinematic viscosity equal to $\nu$. The flow is stationary and incompressible.

The jump condition (\ref{visco}) is applied in this case for cells cut by the solid wall represented by viscosity $\nu_2 \rightarrow \infty$ defining the channel, in practice $\nu_2 = 10^{15 }$. A simple calculation leads to a viscosity value $\nu_m$ which depends only on the viscosity of the fluid moving in the channel, $\nu_m = \mathcal S / \mathcal S_1$ where $\mathcal S_1$ is the portion of surface occupied by this fluid.
The equation of motion is derived from discrete mechanics, but since the problem is stationary, the only physical parameters are $r = dt \: c_l^2$ and the kinematic viscosity $\nu = dt \: c_t^2$. As inertia is zero for this case, the solution can be obtained in a single iteration.
\begin{eqnarray}
\left\{
\begin{array}{llllll}
\displaystyle{ \bm \gamma = - {\mathcal GRAD} \left( \phi^o - r \:  \widetilde{{\mathcal DIV}} \: \widetilde{\mathbf V} \right) + \widetilde{{\mathcal CURL}} \left( \widetilde{\bm \psi^o} - \nu \: {\mathcal CURL} \: \mathbf V \right) +  \mathbf g  } \\  \\
\displaystyle{ \widetilde{\bm \psi^o} - \nu \: {\mathcal CURL} \: \mathbf V \longmapsto \widetilde{\bm \psi^o}}
\end{array}
\right.
\label{poisdis}
\end{eqnarray}
with $\mathbf g = 1 \: \mathbf e_x = {\mathcal GRAD} \: \phi_g =  {\mathcal GRAD} ( - x ) = - \widetilde{{\mathcal CURL}} \: \widetilde{\bm \psi_g} = - \widetilde{{\mathcal CURL}} ( y  )$. The numerical parameter $r = dt \: c_l^2$ makes it possible to maintain a divergence that is as small as necessary, an order of magnitude $\widetilde{{\mathcal DIV}} \: \widetilde{\mathbf V} \approx 1 / r$.
\begin{figure}[!ht]
\begin{center}
\includegraphics[width=7.cm]{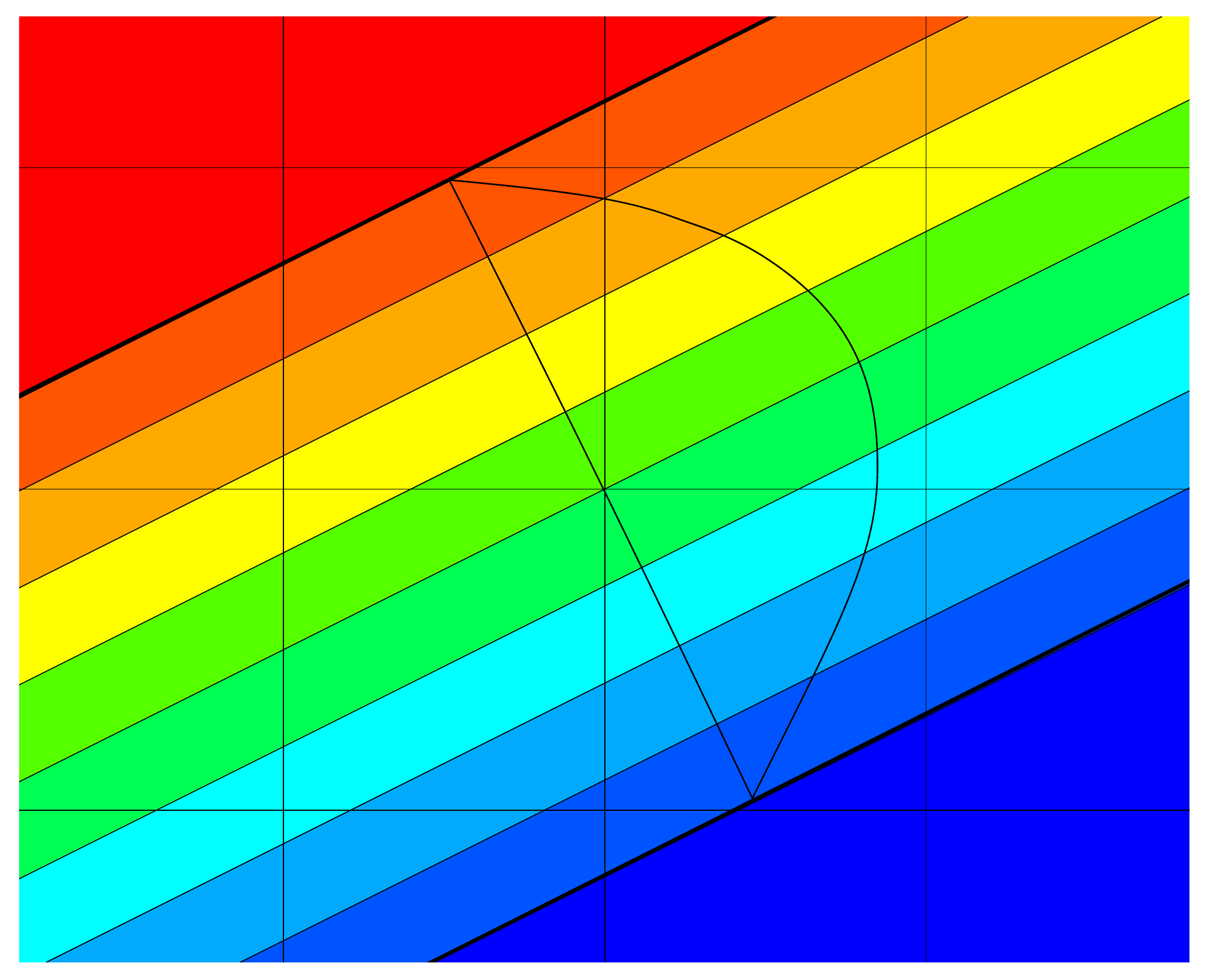}
\caption{\it Poiseuille flow in a planar channel $h = [\pm 0.1342]$ inclined to $\theta = 26.56$° in a domain of dimension $(\pm 1)$; potential vector of acceleration $\widetilde{\bm \psi^o} \in [\pm 2.24]$; the mesh is $8^2$ regular Cartesian cells. The scalar potential $\phi^o$ and the divergence of the velocity $\widetilde{{\mathcal DIV}} \: \widetilde{\mathbf V}$ are null to machine precision. }
\label{traverpois-a}
\end{center}
\end{figure}

\begin{table}[!ht]
\begin{center}
\begin{tabular}{|c|c|c|c|c|}   \hline
    $n^2$    &   Residual $\mathbf V$    &   Residual $\bm \psi^o$   \\ \hline  \hline
  $4^2$      &  $3.30 \: 10^{-16}$     &     $5.55 \: 10^{-16}$      \\ \hline
  $8^2$      &  $4.55 \: 10^{-16}$     &     $1.04 \: 10^{-15}$     \\ \hline
  $16^2$     &  $1.51 \: 10^{-15}$     &     $2.82 \: 10^{-15}$ \\ \hline
\end{tabular}
\caption{\it Poiseuille flow in an inclined channel; residual on the velocity in norm $L_2$ between the theoretical solution projected on the segments of the primal mesh and the numerical solution; residue on the vector potential $\widetilde{\bm \psi^o}$. }
\label{travers}
\end{center}
\end{table}

The solution shown in figure (\ref{traverpois-a}) corresponds to an equal steady-state potential at $\widetilde{\bm \psi^o} = \nu \: {\mathcal CURL} \: \mathbf V$ and a velocity equal to $\mathbf V \cdot \mathbf e_u = y^2 + a \: y + b$ where $\mathbf e_u$ is the axial vector of the inclined channel.
The figure schematically represents the parabolic profile of the velocity and of the current lines between the two solid walls.

Table (\ref{travers}) gives the residuals between the theoretical solution and the numerical simulations on the velocity and the vector potential for three different meshes, thus showing the independence of the spatial approximation for the problem posed.

\textcolor{blue}{\subsection{Two-Phase Poiseuille Flow} }

For this example, the two fluids have finite viscosity of $\nu_1$ and $\nu_2$. Since the variations in potentials are linear, the homogenized viscosity has a simple form:
\begin{eqnarray}
\displaystyle{  \nu_m = \left( \frac{1}{\nu_1} \: \frac{\mathcal S_1}{\mathcal S} + \frac{1}{\nu_2} \: \frac{\mathcal S_2}{\mathcal S}  \right)^{-1} } 
\label{vispois}
\end{eqnarray}
As the geometry of the channel is fixed in time, it is easy to calculate $\nu_m$ for each cell. In steady state, the density does not affect the solution.

Consider a duct of height $h = 1$ bounded by two horizontal surfaces of zero velocity. The lower kinematic viscosity fluid $\nu_1 = 1$ and the higher viscosity fluid $\nu_2 = 10$ occupy the space respectively between $y \in [0, y_0]$ and $y \in [y_0, 1]$. The fluid is set in motion by a pressure gradient or vector potential gradient in discrete mechanics, such that $\widetilde{{\mathcal CURL}} \:  \widetilde{\bm \psi^o} = 1.2$. The interface is positioned at value $y_0 = 0.6$ and does not correspond to a particular value associated with the mesh.
The initial velocity and scalar potential values are zero. The solution can be obtained in a single iteration since inertia is zero for the problem posed.
\begin{figure}[!ht]
\begin{center}
\includegraphics[width=8.cm]{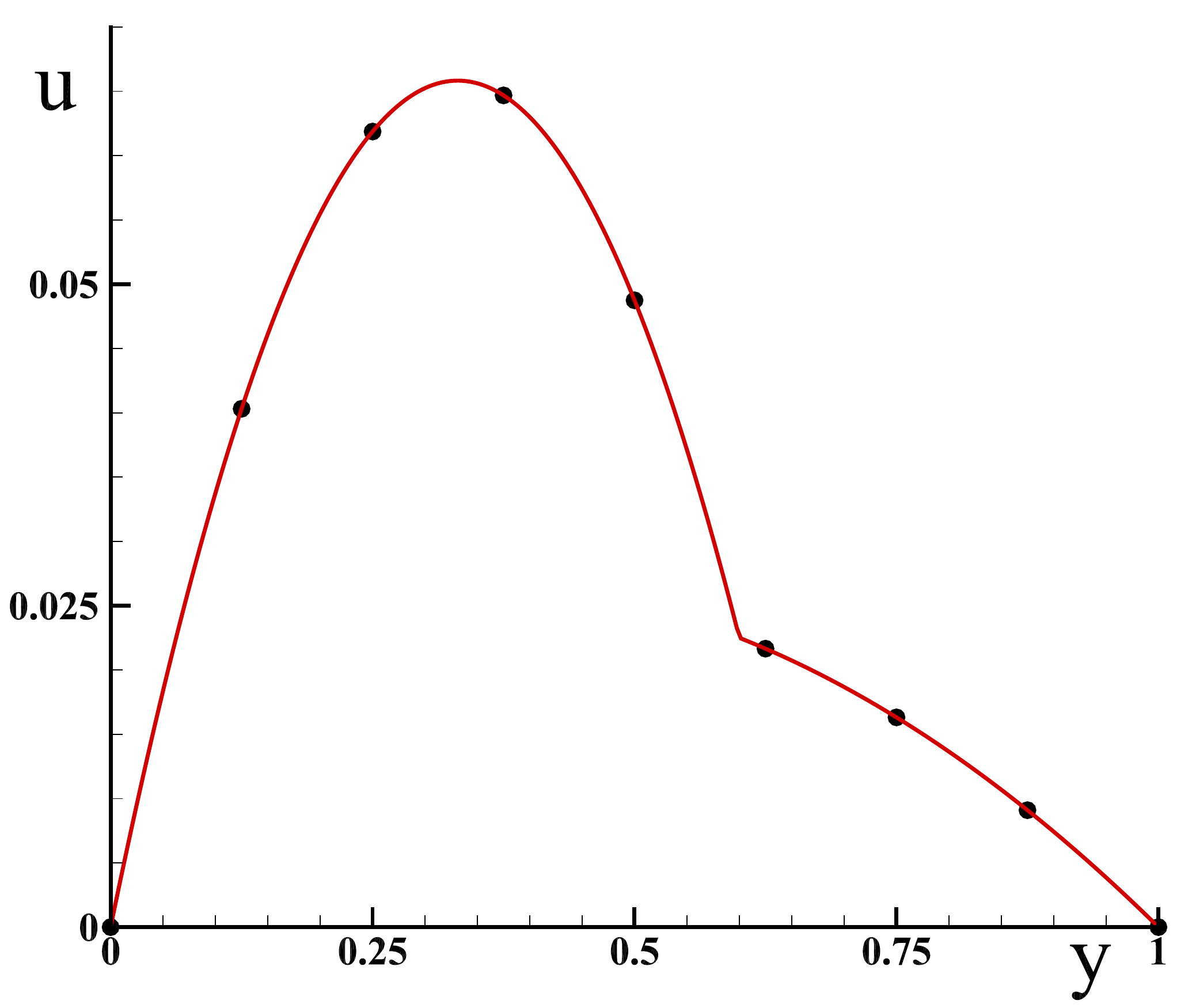}
\caption{\it Two-phase Poiseuille flow of height $h = 1$ where the interface is positioned in $y_0 = 0.6$. The viscosity of the fluid at the bottom is $\nu_1 = 1$ and that of the higher fluid is $\nu_2 = 10$; the mesh is $8^2$ cells. Comparison between the analytical solution (line) and the numerical values (points); the divergence of the velocity $\widetilde{{\mathcal DIV}} \widetilde{\mathbf V}$ and the scalar potential $\phi^o$ are null to machine precision.}
\label{poisdiph}
\end{center}
\end{figure}

The condition at the boundary between the two fluids, whose velocity fields are $\mathbf V_1$ and $\mathbf V_2$, is given implicitly by the discrete formulation of the equation of motion. The residue in norm $L_2$ on the velocity and on the vector potential is given by table (\ref{poisdiphr}):
\begin{table}[!ht]
\begin{center}
\begin{tabular}{|c|c|c|c|c|}   \hline
    $n^2$        &   Residual $\mathbf V$    &   Residual $\widetilde{\bm \psi^o}$   \\ \hline  \hline
  $8^2 $     &  $6.45 \: 10^{-18}$     &     $1.27 \: 10^{-16}$ \\ \hline
\end{tabular}
\caption{\it Two-phase Poiseuille flow in planar channel; residual on the velocity in norm $L_2$ between the theoretical solution projected to the barycenters of the primal mesh and the numerical solution and residue on the vector potential $\widetilde{\bm \psi^o}$. }
\label{poisdiphr}
\end{center}
\end{table}

The solution is accurate to machine precision for all quantities, regardless of the number of cells adopted.
The test case of the horizontal planar channel is taken up in a mesh based on triangles. With triangular meshes made with any software, several imperfections usually appear; for example the orthogonality of the segments of the primal topology and those connecting the barycenters is not assured. If the objective is to achieve machine precision on the numerical solution, then it is necessary that all the metrics be evaluated with the machine precision, for example the position of the points must be given with $15$ significant digits in double precision.

\begin{figure}[!ht]
\begin{center}
\includegraphics[width=8.cm]{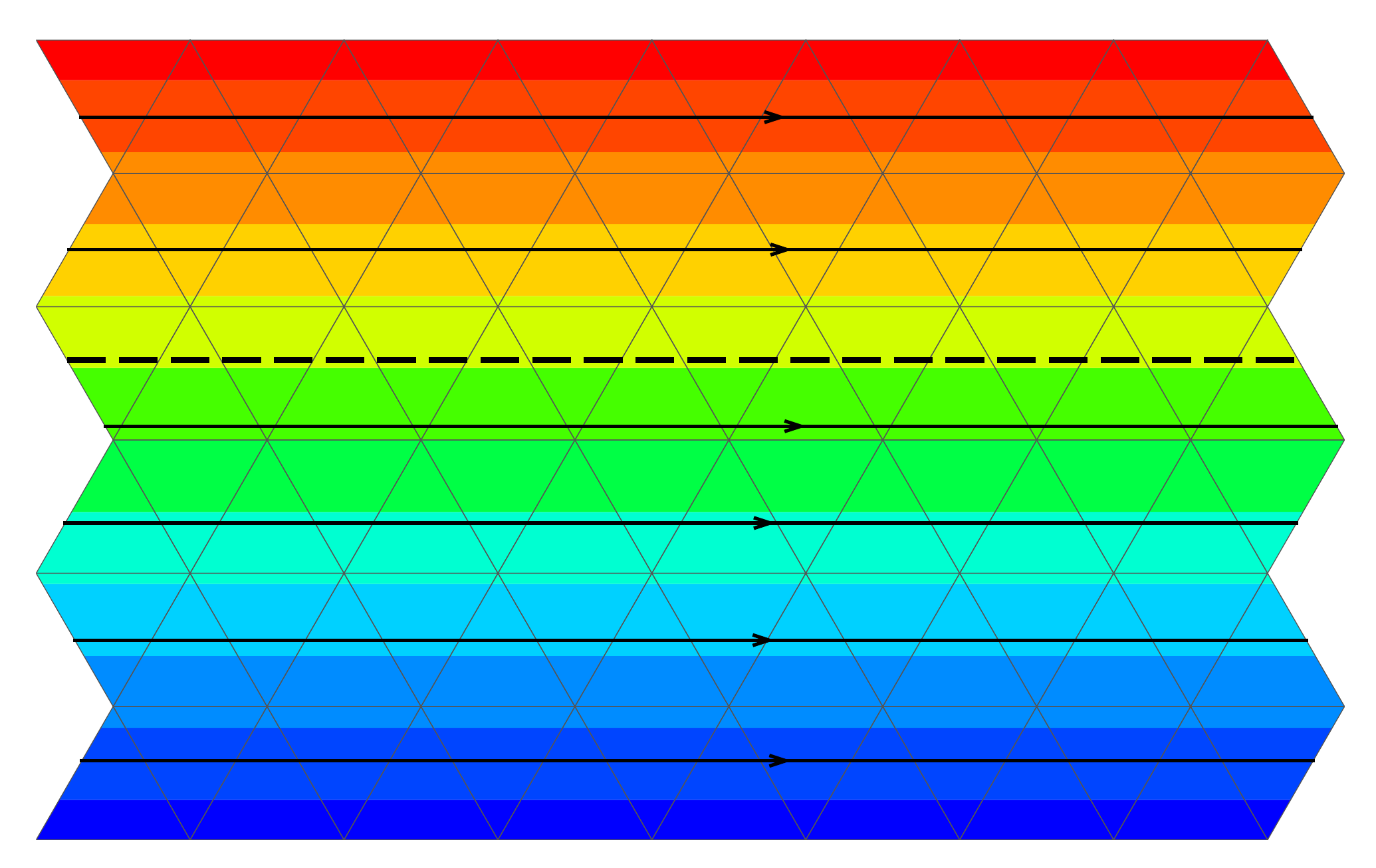}
\caption{\it Two-phase Poiseuille flow of height $h = 1$ where the interface is positioned in $y_0 = 0.6$. The viscosity of the fluid at the bottom is $\nu_1 = 1$ and that of the higher fluid is $\nu_2 = 10$; the mesh is $16 \times 6$ cells. Vector potential field $\widetilde{\bm \psi^o}$ and some current lines; the values of the divergence of the velocity $\widetilde{{\mathcal DIV}} \widetilde{\mathbf V}$ and the scalar potential $\phi^o$ are null to the machine precision.}
\label{poisdiphex}
\end{center}
\end{figure}

Generally, the results are in the order of two, but the absolute error in the vicinity of the interface between the two fluids is very important, whatever the interpolation adopted for the viscosity. Since the theoretical solution corresponds to two polynomials of degree equal to two, it is possible to find an exact numerical solution (figure \ref{poisdiphex}).
 
The height channel $h = 1$ is filled with two fluids of viscosity $\nu_1$ and $\nu_2$ with an interface $\Sigma$ located at $y_0 = 0.6$. The equilateral triangle mesh contains $16 \times 6$ cells.

The numerical solution is represented in figure (\ref{poisdiphex}) which shows the mesh used, the vector potential field and some current lines. The interface $\Sigma$ is set to $y_0 = 0.6$ and the stream lines are unaffected by changes in viscosity.
\begin{table}[!ht]
\begin{center}
\begin{tabular}{|c|c|c|c|c|}   \hline
    $n^2$        &   Residual $\mathbf V$    &   Residual $\widetilde{\bm \psi^o}$   \\ \hline  \hline
  $8^2 $     &  $4.02 \: 10^{-17}$     &     $1.77 \: 10^{-16}$ \\ \hline
\end{tabular}
\caption{\it Two-phase Poiseuille flow in planar channel for a uniform mesh based triangles; residual on the velocity in norm $L_2$ between the theoretical solution projected to the barycenters of the primal mesh and the numerical solution and residuals on the vector potential $\widetilde{\bm \psi^o}$. }
\label{poisdiphrex}
\end{center}
\end{table}

Table (\ref{poisdiphrex}) gives the residue of the numerical solution on the vector potential and velocity. The viscosity ratio here is equal to $ 10 $ but it is possible to set ratios of $0$ (for a non-conforming free surface $\mathbf V \cdot \mathbf n = 0$) to infinity (for a non-conforming solid wall $\mathbf V = 0$). This formulation makes it possible to process 3D interfaces without modification to address complex geometry problems with non-conforming meshes.

\textcolor{blue}{\subsection{Hadamard-Rybczynski fluid flow} }

The Hadamard-Rybczynski solution \cite{Had11, Tav20} corresponds to an incompressible, stationary non-inertial flow of a bubble in a fluid of different density under the effect of the Archimedean force. In fact, as the densities are constant in each of the fluids, the source term to be integrated into the equation of motion is constant. Since the external solution is in $1 / r^2$, the numerical solution can only be at the order of the scheme and the absolute error then depends on it. On the other hand, the internal solution corresponds to a polynomial of degree equal to two and it is then possible to obtain the theoretical solution to machine precision whatever the spatial approximation.

The solution is sought from the equation of motion (\ref{discrete}) solved in the reference frame of the sphere. The separation of the variables was used by Hadamard and Rybczynski to obtain the solution:
\begin{eqnarray}
\left\{
\begin{array}{llllll}
\displaystyle{ \mathbf V = K \left[ \left( 1 +  r^2 \right) \: \cos \theta \: \mathbf e_r -  \left( 1 + 2 \: r^2 \right)  \: \sin \theta \: \mathbf e_{\theta} \right] } \\  \\
\displaystyle{   \bm \psi = - 4 \: K \: r \: \sin \theta \: \mathbf e_{\varphi} }
\end{array}
\right.
\label{hadam}
\end{eqnarray}
 with $r = \sqrt{x^2 + z^2}$ ($\mathbf e_r = \mathbf e_x + \mathbf e_z$) and $K = g \: R^2 \: (\rho_e - \rho) / 6 \: (\mu_e + 3 \: \mu / 2)$ where $R$ is the radius of the sphere, $\mu$ and $\rho$ the viscosity and density of the inner fluid and $\mu_e$ and $\rho_e$ those of the exterior fluid. For the simulation, the density and viscosity dependent parameter is chosen equal to $K = 1$. The source term of the system (\ref{discrete}) is written $\bm h = - {\mathcal GRAD} \: \phi_g + \widetilde{{\mathcal CURL}} \: \widetilde{\bm \psi_g} = - \nabla ( -r) + \nabla \times ( y \:\: \mathbf e_{\varphi} )$.

\vspace{2.mm}
\paragraph{\textcolor{blue}{2D-axisymmetric solution: } }

The methodology used is based on an unstructured mesh not conforming to the unit circle representing the spherical bubble (treated axisymmetrically). The equilateral triangle mesh requires the triangles cut by the circular interface $\Sigma$ to be treated and the non-intersected outer triangles to be excluded. A series of differential geometry algorithms allows calculation of the area of polygon portions with one side curved. The reconstitution of the unit circle makes it possible to calculate its area $\mathcal A = \pi \: R^2$ to machine precision.
From the information on the proportion of the area of the triangle belonging to the domain, it is possible to calculate the viscosity of each of the cells cut by the interface. 
\begin{figure}[!ht]
\begin{center}
\includegraphics[width=5.cm]{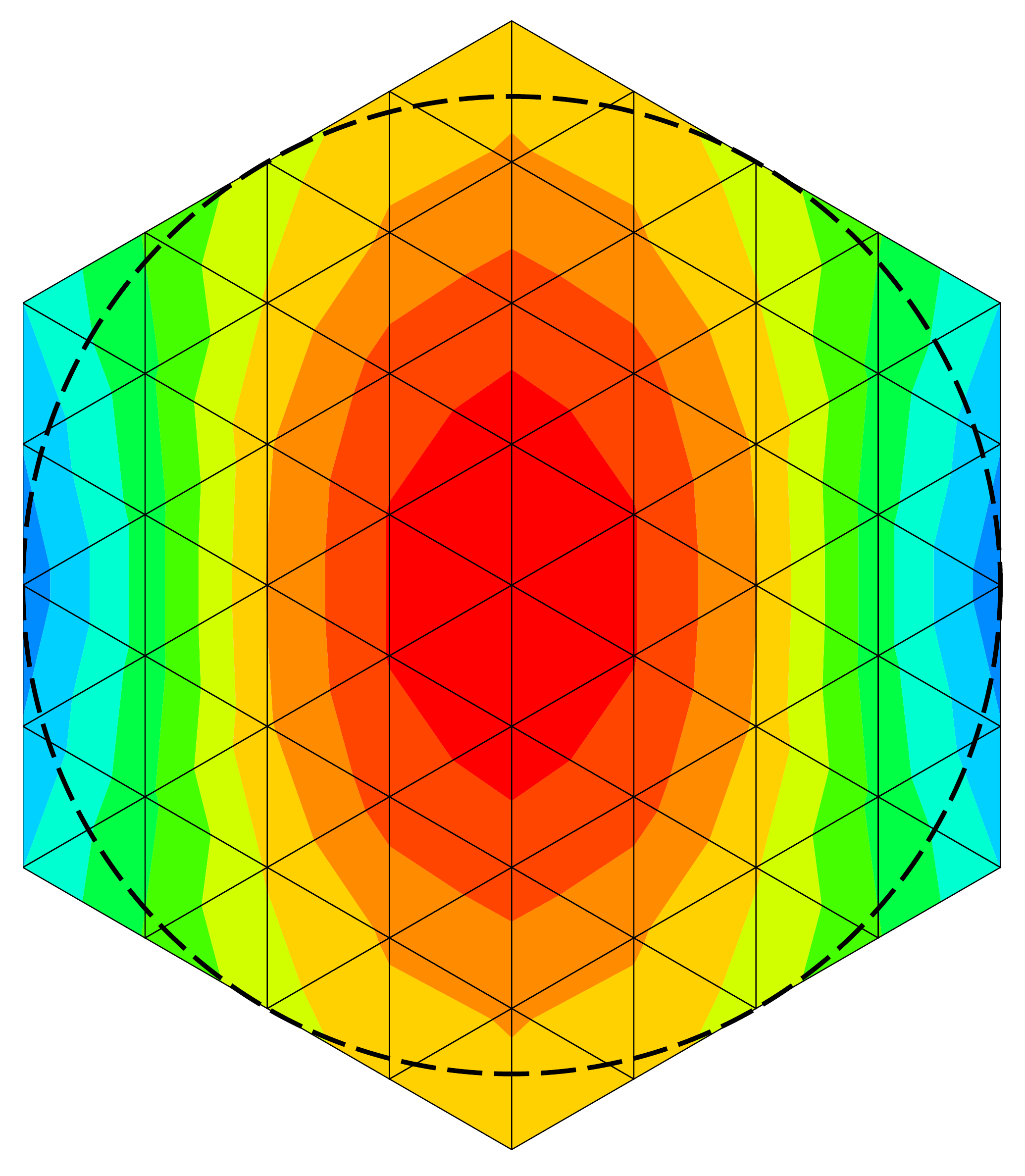}
\includegraphics[width=5.cm]{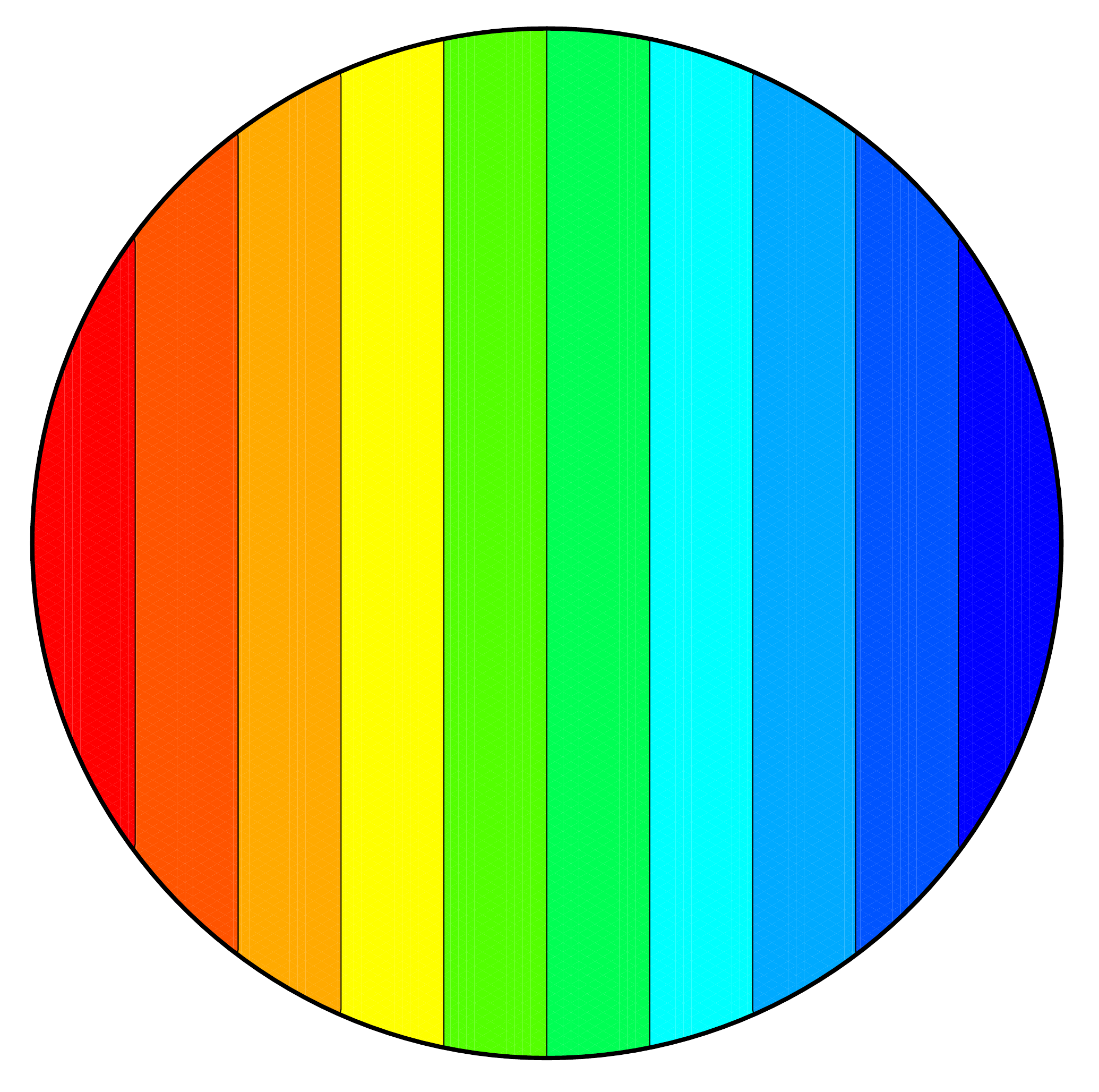}
\includegraphics[width=5.cm]{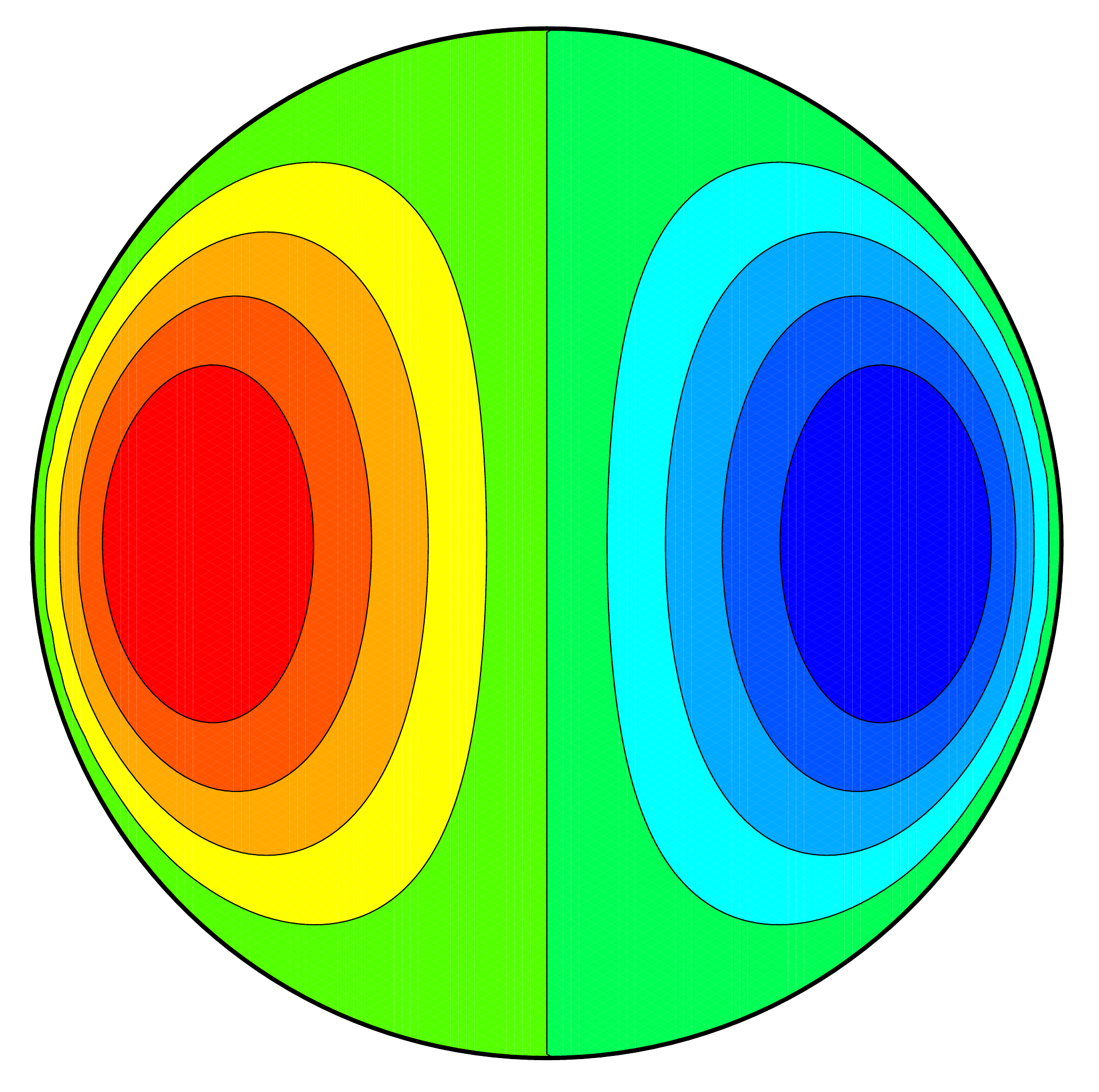}
\caption{\it Hadamard-Rybczynski's solution inside the sphere; left vertical velocity for a mesh of $n = 96$ elements; center and right, vector potential $\bm \psi$ and current lines for a mesh of $n = 22608$ equilateral triangles. The solution is exact and the divergence is zero.}
\label{hadamard-2d}
\end{center}
\end{figure}

The source term and incompressibility are sufficient to define this test case completely. The separation of the variables used by Hadamard to obtain the solution is proof that the effects of compression and shear are disjoint.
Figure (\ref{hadamard-2d}) shows the vertical velocity in the bubble and the vector potential $\bm \psi$, exact to machine precision regardless of the spatial approximation. Values outside the circle have no meaning. Since the current function is not a primitive variable of discrete mechanics, it is recalculated on the points by solving a Poisson equation; the current function is indeed only one component of the vector potential of the velocity, whereas $\bm \psi$ is the vector potential of the acceleration.
The velocity components and vector potential $\bm \psi$ obtained by numerical simulation are accurate to machine precision regardless of the spatial approximation. The solution is represented in figure (\ref{hadamard-2d}). It is obtained in a single step and a few conjugate gradient iterations.

\paragraph{\textcolor{blue}{3D-solution:} }

The solution to this problem in a three-dimensional space is treated using a structured mesh based on regular hexahedrons, where the outer facets not concerned by the flow in the bubble are eliminated {\ it a priori}. As regards the principle of the methodology, nothing is changed from the 2D-axisymmetric approach and the same procedure is applied whatever the orientation of the facets. The trace of the sphere of unit radius on the facets makes it possible to calculate the areas $\mathcal S_1$ of the parts of the spherical domain. The boundary conditions on the sphere naturally correspond to a slip of the fluid.
\begin{figure}[!ht]
\begin{center}
\includegraphics[width=6.cm]{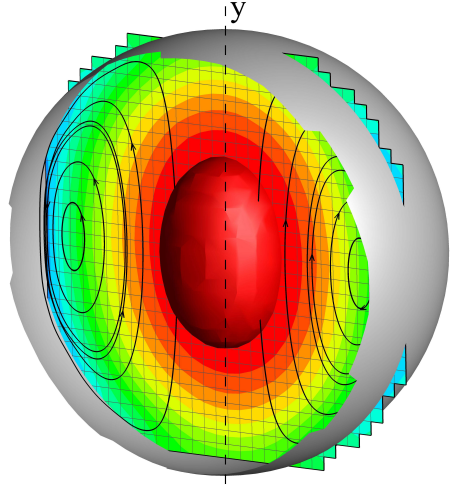}
\caption{\it Hadamard-Rybczynski 3D solution inside the sphere; the mesh is composed of $32^3$ hexahedral cells. Stream lines and equipotentials show that the velocity field and the vector potential $\bm \psi$ are axi-symmetric. The stationary solution is exact and the divergence is zero.}
\label{hadamard-3d}
\end{center}
\end{figure}

Figure (\ref{hadamard-3d}) represents the solution obtained directly with system (\ref{discrete}) graphically. The solution obtained on the velocity field is exact and the vector potential $\bm \psi$ is strictly linear per surface. This example treated with a mesh conforming to the unstructured sphere inevitably leads to errors due to the tessellation of the sphere surface and the polyhedral volume mesh. The choice of a Cartesian structured mesh is of considerable interest for various reasons and the application of these jump treatment techniques makes it possible to solve one of its difficulties, consisting in taking account of the complex geometry. 

\textcolor{blue}{\subsection{Droplet in equilibrium under capillary acceleration} }

The incompressible equilibrium of a droplet subjected to capillary acceleration has absolutely no dependence on the viscosity or density of the fluids. The pressure difference between fluids is given by Laplace's law $ \Delta p = \gamma \: \kappa $, or, in discrete mechanics, by $\Delta \phi^o = \sigma \: \kappa$. Indeed, an incompressible movement, linked to an infinite celerity of sound $c \rightarrow \infty$, leads to an instantaneous equilibrium. The equation (\ref{discrete}) in the presence of a capillary force differs significantly from the Navier-Stokes equation; in particular the conventional couplings between velocity and pressure are replaced by an integration of the incompressibility constraint in the equation itself. This problem does not generate {\it a priori} any rotational one usually interpreted as spurious currents. The equation of motion is in this case:
\begin{eqnarray}
\displaystyle{ \bm \gamma = - {\mathcal GRAD} \left( \phi^o - r \:  \widetilde{{\mathcal DIV}} \: \widetilde{\mathbf V} - \sigma \: \kappa \: \xi \right)   } 
\label{capill}
\end{eqnarray}

\begin{figure}[!ht]
\begin{center}
\includegraphics[width=5.cm]{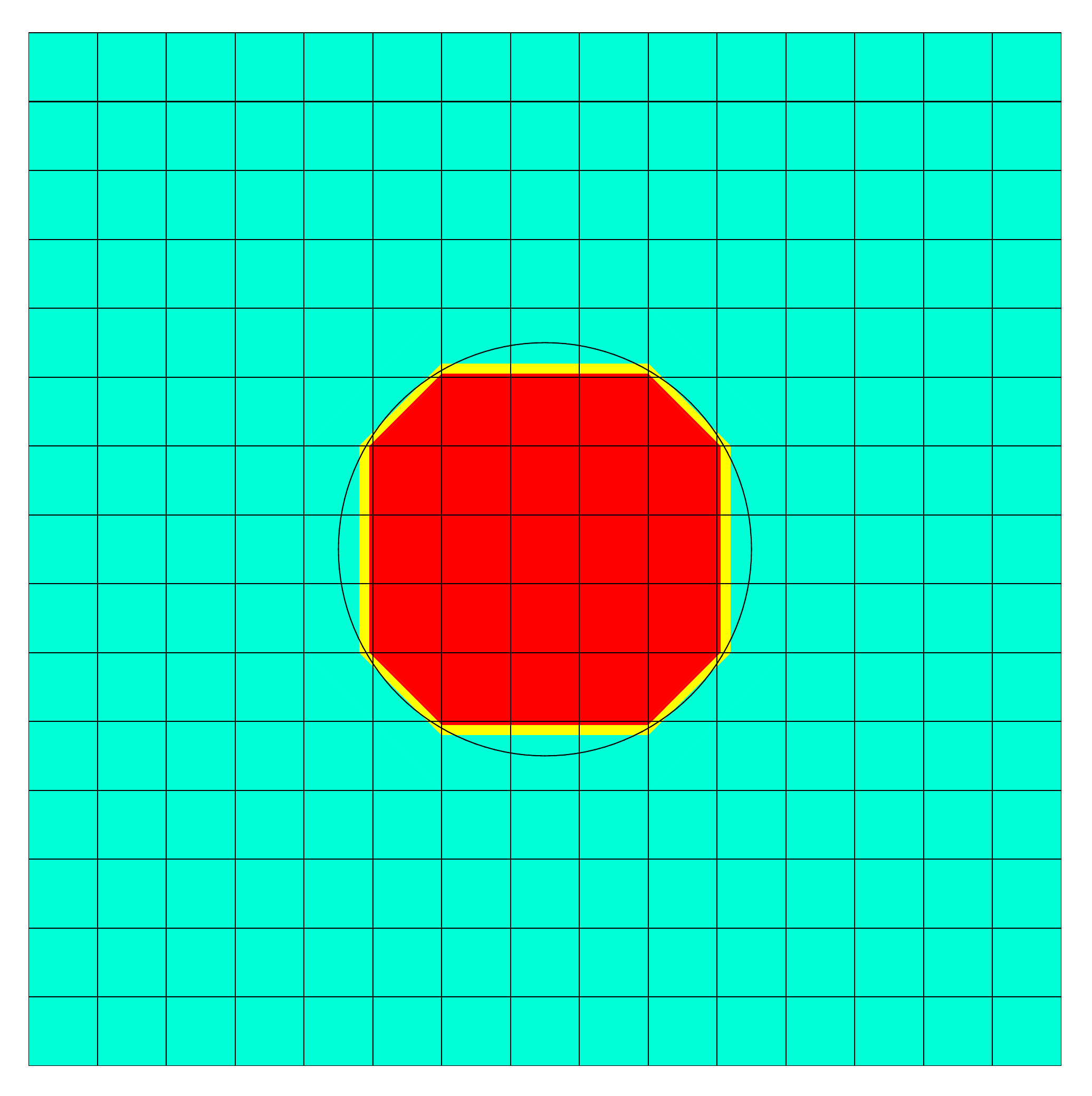}
\includegraphics[width=5.cm]{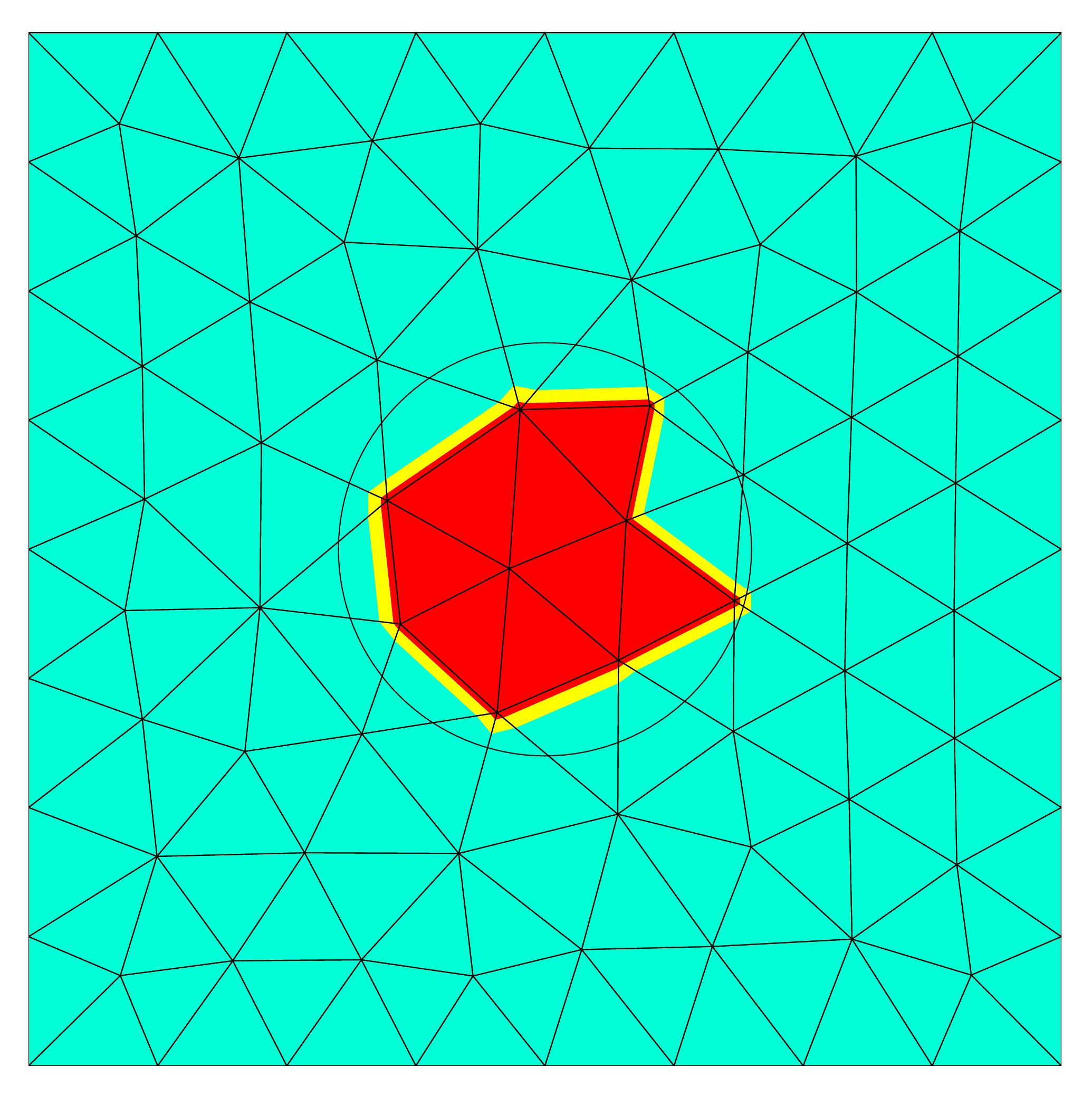}
\includegraphics[width=5.5cm]{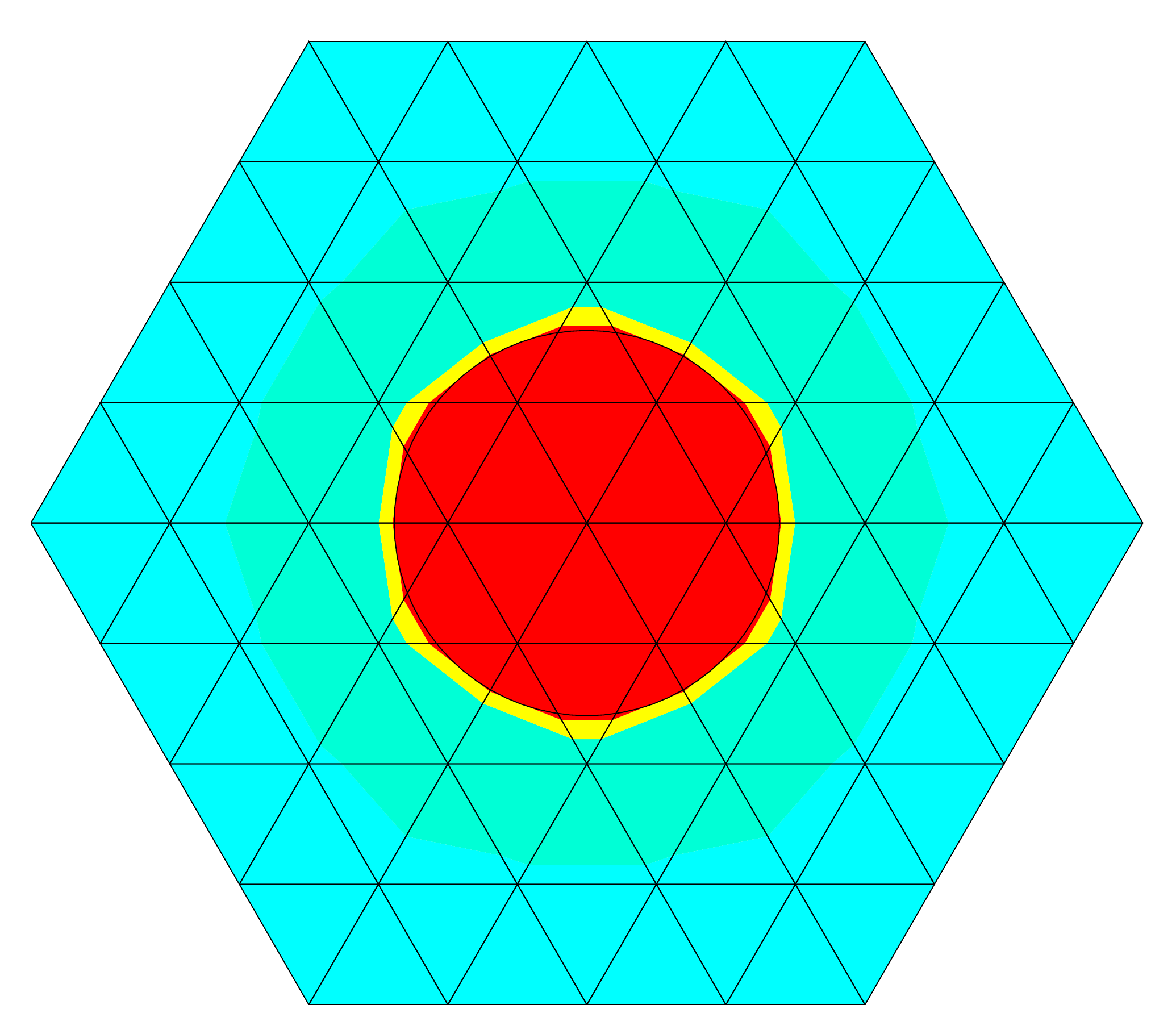}
\caption{\it Equilibrium of a droplet under a capillary effect  simulated using meshes based on Cartesian quadrangles, triangles and equilateral triangles; the potential difference on either side of the interface is exact.}
\label{hexaplace}
\end{center}
\end{figure}

The steady solution of this problem is obtained explicitly by considering the medium as incompressible reducing the equation (\ref{capill}) to $- {\mathcal GRAD} \left( \phi^o  - \sigma \: \kappa \: \xi \right) = 0$; the two terms being gradients the scalar potential of the acceleration can be obtained explicitly by calculating the solution, up to a constant, from one point to another by the relation:
\begin{eqnarray}
\displaystyle{ \phi_b^o = \phi_a^o - \int_a^b \: \nabla \left( \sigma \: \kappa \: \xi \right) \cdot \mathbf t \: dl } 
\label{remonte}
\end{eqnarray}
where $\xi$ is the phase function equal zero or one.
This solution is represented in figure (\ref{hexaplace}) for three meshes, a Cartesian mesh, another an unstructured mesh of triangles and a third on equilateral triangles. 

The simulation with all terms of equation (\ref{discrete}) is performed from null initial pressure and velocity fields.
When the quantity $dt \: c^2$ is very large, the divergence becomes very small and the solution of $- {\mathcal GRAD} \: \phi^o  + {\mathcal GRAD} \left(\sigma \: \kappa \: \xi \right) = 0$ is obtained instantly.
 A time step of $ dt = 10^{12} \: s$ allows the steady solution to be reached instantaneously where $\mathbf V = 0$ and $\phi^o = \sigma \kappa$. The accuracy of the solution is order of magnitude of machine precision; the difference between analytical and numerical solution it is kept smaller than $10^{-16}$ after 2 iterations. 
The physical model presented is suitable for solving various problems in mechanics. For a fixed intrinsic celerity, the nature of the flow depends on the observation time scale $dt$; for example, the movements of water, an essentially incompressible liquid $ (c = 1500 \: m \: s^{-1})$, cause acoustic waves to appear with weak time constants. Excluding microscopic interactions, the model (\ref{discrete}) is representative of physical phenomena at all scales of time and space.
\begin{figure}[!ht]
\begin{center}
\includegraphics[width=7.cm]{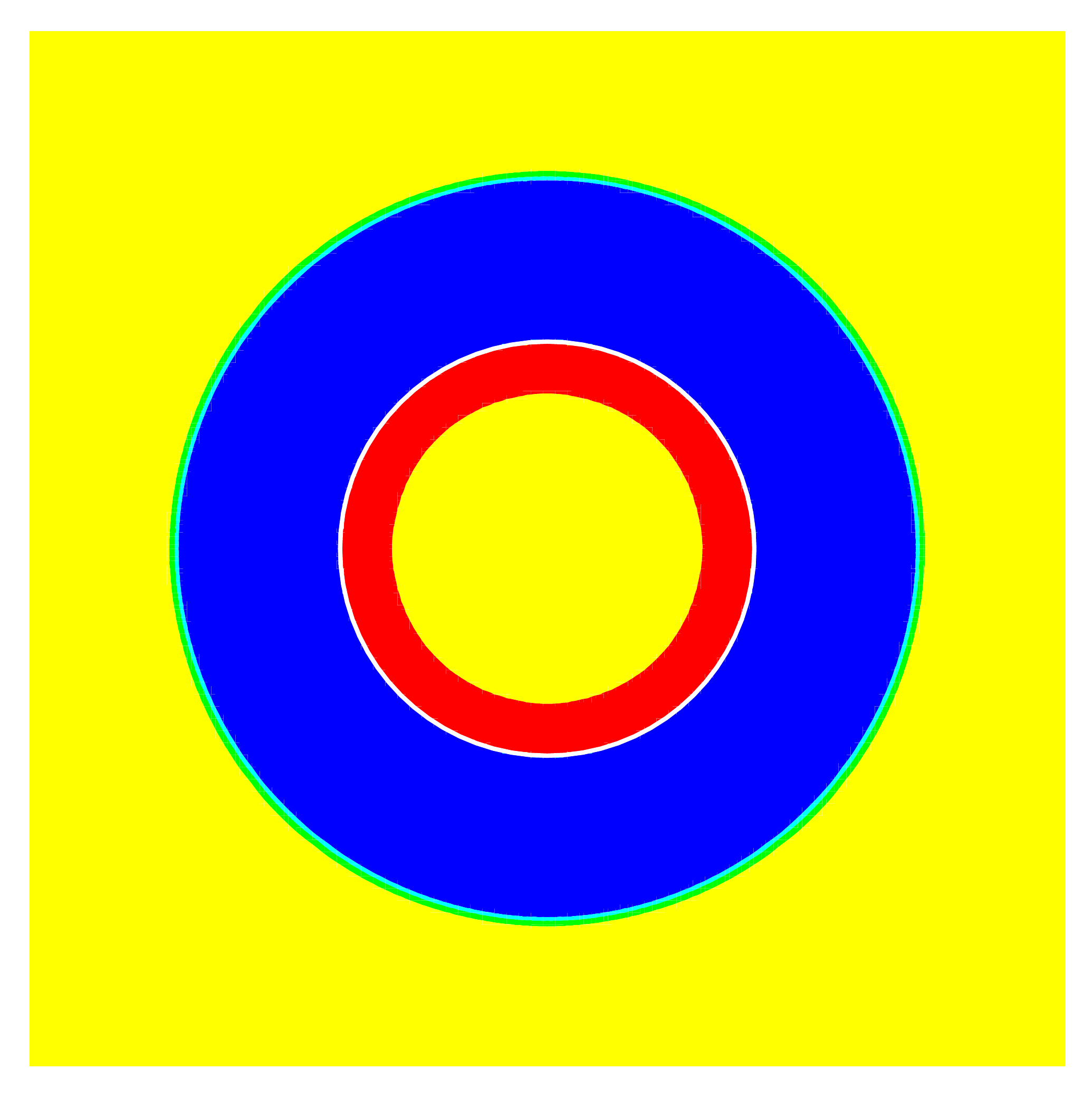}
\caption{\it Droplet of radius $R = 2 \: 10^{-3} \: m$ in a square cavity filled with two inviscid fluids; scalar potential field; Cartesian mesh of $256^2$ cells, $dt = 10^{-7} \: s$, $t = 5 \: 10^{-5} \: s$ with $c = 10^3 \: m \: s^{-1}$ outside and $c = \sqrt {10^5} \: m  \:s^{-1}$ inside; the outline of the drop is represented by a white circle.   } 
\label{dropletshock}
\end{center}
\end{figure}

The model (\ref{discrete}) is full-compressible and the behavior of the solutions is closely associated with the longitudinal and transverse waves which propagate in the media. For the two fluids of the Laplace test case, the transverse waves attenuate very quickly but there remain the longitudinal waves which propagate at the celerity of sound $c = 1 / \sqrt {\rho \: \chi_T}$, where $\chi_T$ is the compressibility coefficient. Thus, if the scalar potential is zero at the initial moment, the capillary acceleration ${\mathcal GRAD} \left(\sigma \: \kappa \: \xi \right)$ causes a radial collapse of the drop on itself, compensated by the quasi-incompressibility of the fluid; as the fluids have finite celerities the stationary solution is not reached instantly if the time steps $ dt $ are lower than $ R / c $. Figure (\ref{dropletshock}) shows the solution at time $t = 5 \: 10^{-5} \: s$ obtained with a time-step $dt = 10^{-7} \: s$ for two non-viscous fluids. 

As the two fluids have different properties, the radial expansion wave in the external fluid and the compression wave in the drop propagate at different celerities, since the square cavity is closed. Continuing the simulation leads to a succession of compressions and detents which interact with the walls of the cavity. For inviscid fluids, the kinetic energy should be conserved over time provided that the phenomenon is simulated with sufficient temporal precision.
\begin{figure}[!ht]
\begin{center}
\includegraphics[width=10.cm]{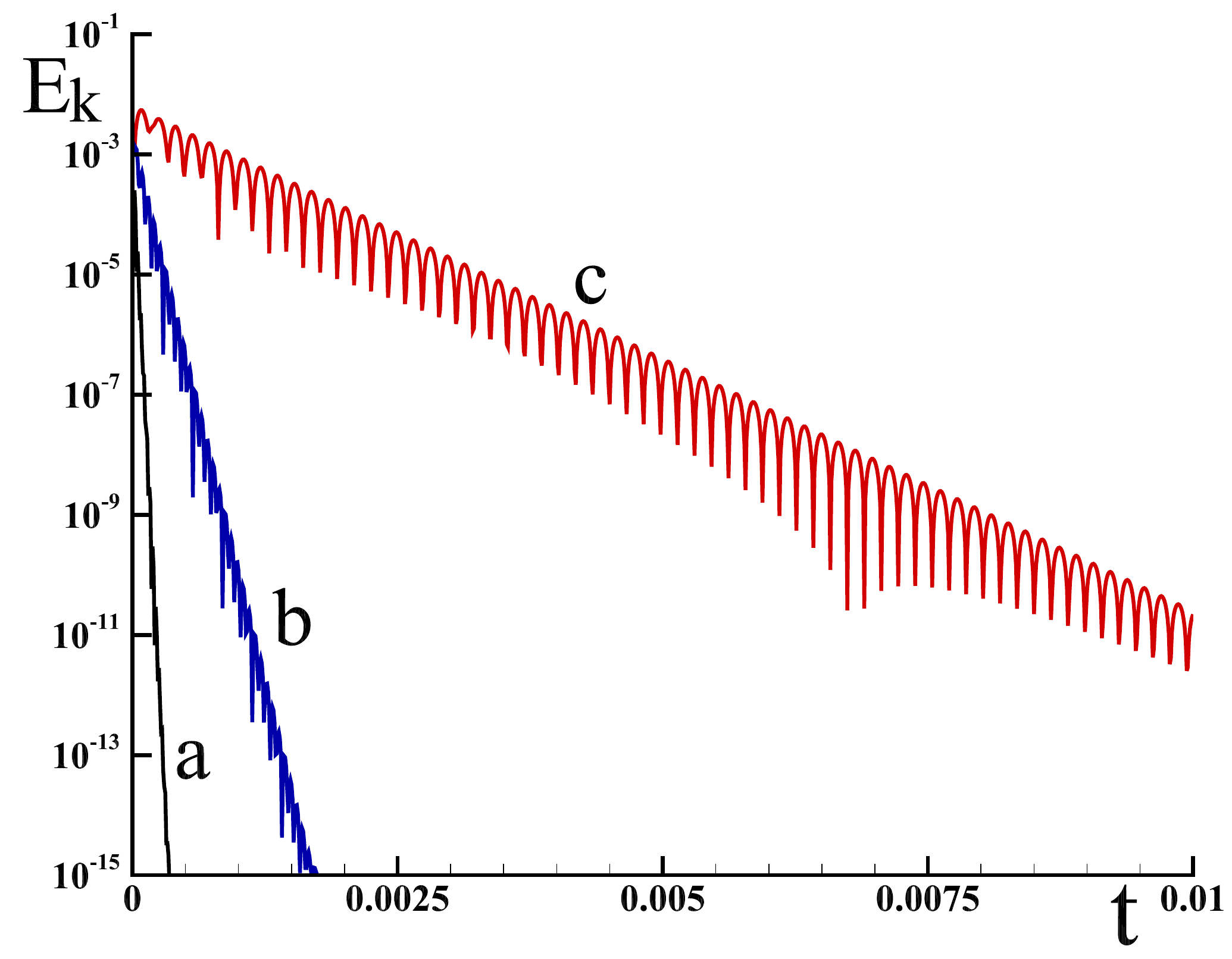}
\caption{\it Kinetic energy in a droplet of radius $R = 2 \: 10^{-3} \: m$ in a square cavity filled with  inviscid fluids; norm of velocity,  mesh of $64^2$ and time-step  $dt = 10^{-5} \: s$, with various celerities of sound (a) $c=10^5\: m \: s^{-1}$, (b)  $c=10^4 \: m \: s^{-1}$, (c)  $c=10^3 \: m \: s^{-1}$}
\label{dropletconv}
\end{center}
\end{figure}

Depending on the values of the celerity of sound, the steady solution of capillary equilibrium is obtained more or less quickly. The oscillations of the velocity norm represented in figure (\ref{dropletconv}) are absolutely not parasitic currents but reflect the temporal variations of the potential $\phi$ of the compressible movement. The incompressible motion $c \rightarrow \infty$ must lead to an instantaneous capillary equilibrium and the incompressible model causes the instantaneous extinction of all acoustic waves. 
 
We can see that (i) the solution does not depend on the main mesh, (ii) the curvature obtained with differential geometry routine is exact , (iii) the oscillations of the velocity norm are independent of the viscosity, (iv) the solution is independent of density.


The interpretation of the results of the dynamic and static equilibrium of a drop subjected to a capillary pressure jump differs from that generally given for simulations carried out in the context of a continuous medium for an incompressible movement. The Capillary $Ca$, Laplace $La$ or Ohnesorge $Oh$ numbers representative of this problem involve the viscosity of fluids and their density.

The Continuum Surface Force (CSF) model generally introduced in continuum mechanics to describe the capillary term as $\gamma \, \kappa \, \delta_{\Sigma} \, \mathbf n = \gamma \, \kappa \, \nabla c$ has no reason to be a gradient, in general. The spurious currents observed by many authors are due to curvature calculation errors which generate vorticity whose intensity depends on the numerical methodology used.

\textcolor{blue}{\subsection{Free oscillation of viscous droplet} }

The oscillations of a drop subjected to capillary accelerations are an opportunity to highlight the originality which consists in removing the density of the quantities present in the incompressible Navier-Stokes equation. Indeed, the two physical quantities present in the discrete equation (\ref{discrete}) are the celerities $c_l = 1 / \sqrt {\rho \: \chi_T}$ and $ c_t = \sqrt {\mu / \rho }$ which show three physical quantities, $\rho $, $\mu $ as well as $\chi_T$ whose measurement is generally carried out indirectly by that of the celerity. For time constants greater than $\tau \approx 10^{-10} \: s$, the grouping $dt \: c_t^2$ must be replaced by the kinematic viscosity $ \nu $ for homogeneous and isotropic fluids. In addition to the reduction of physical quantities, one can observe with  table (\ref{propert}) that, in the case of a two-phase water-air flow, the ratios of the selected quantities are much lower than those adopted conventionally. 
\begin{table}[!ht]
\begin{center}
\begin{tabular}{|c|c|c|c|c|c|c|}   \hline
     & $\rho$ &  $\chi_T$        & $\mu$       & $c_l $      &  $\sqrt{\nu}$    \\ \hline  \hline
 water & $1000$    &   $10^{-9}$ &   $10^{-3}$ &   $1000$    &  $10^{-3} $  \\ \hline
 air   & $1$       &   $10^{-5}$ &   $10^{-5}$ &   $316$     &  $3.16 \: 10^{-3}$   \\ \hline
 ratio & $10^{3}$  &   $10^{-4}$ &   $10^{-2}$ &   $3.16$    &  $0.316 $  \\ \hline
\end{tabular}
\caption{\it Physical parameters of two fluids close to those of water and air. }
\label{propert}
\end{center}
\end{table}

The historical work of Lord Rayleigh \cite{Ray79b, Kel90} on the oscillations of drops and bubbles served to define the frequency of infinitesimal-amplitude oscillations of incompressible liquid drops about the spherical shape in vacuum. More recent studies, including \cite{Lam93}, \cite{Cha59}, \cite{Pro80}, \cite{Pro12}, have extended the field of investigation to dissipative effects using linear theories. The problem of the oscillation of a droplet in nonlinear theory for a small Ohnesorge number has been addressed recently \cite{Plu20} using the potential flow assumption to reduce the corresponding free boundary problem formulated on a time-dependent domain into a nonlinear system of integro-differential equations.  
For two-dimensional planar drops oscillating about a circle, the frequency of the oscillations is given by:
\begin{eqnarray}
\displaystyle{ f^2_{th} = m \: (m^2 - 1 )\: \frac{\gamma}{\rho \: R^3} } 
\label{frequency}
\end{eqnarray}
where $m$ is the mode number and the density $\rho$ is that of the drop fluid. The frequency is independent of the viscosity but the attenuation of the oscillations depends on it.

The test case is chosen so as to show that the formulation of the jumps in density, viscosity and capillary pressure, as well as the numerical methodology implemented, do not affect the properties of spatial convergence of the system (\ref{discrete}). Consider a circular drop of radius $R = 2 \: 10^{-3} \: m$ deformed into an ellipse with a radius ratio equal to $R_x / R_y = 1.5$ at the center of a square of dimension $R = 10^{-2} \: m$. The properties of the drop recalled in table (\ref{propert}) are very close to those of water, and the physical characteristics of the gas which surrounds it correspond substantially to those of air. The surface tension fixed at  $\gamma = 2 \: 10^{-1} \: kg \: s^{-2}$ or $\sigma = 2 \: 10^{-4} \: m^3 \: s^{-2}$ is chosen small enough to avoid confusing the capillary effects and the acoustic waves of the oscillations of the ellipse over time.

\begin{figure}[!ht]
\begin{center}
\includegraphics[width=5.cm,height=5.cm]{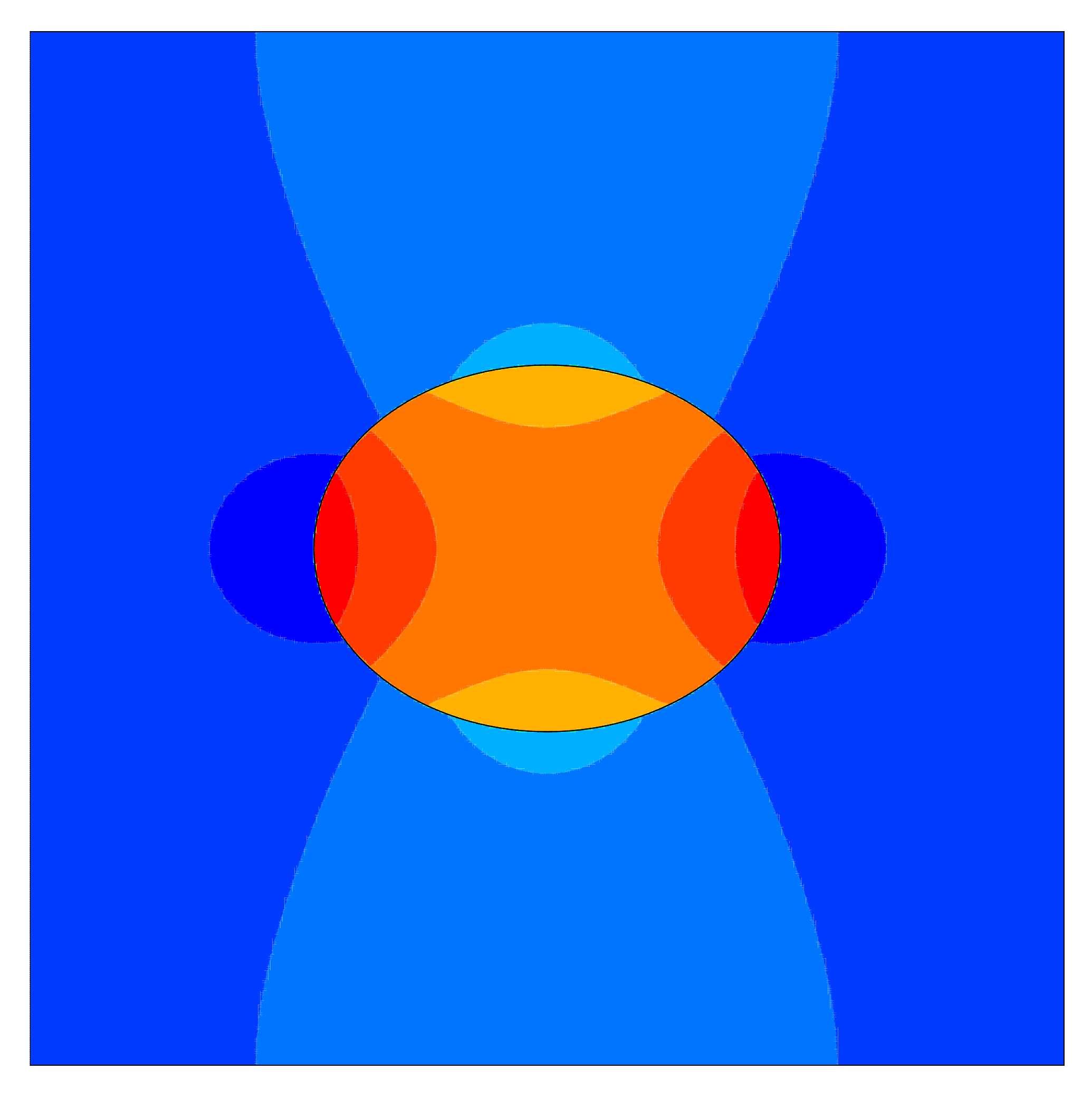}
\includegraphics[width=5.cm,height=5.cm]{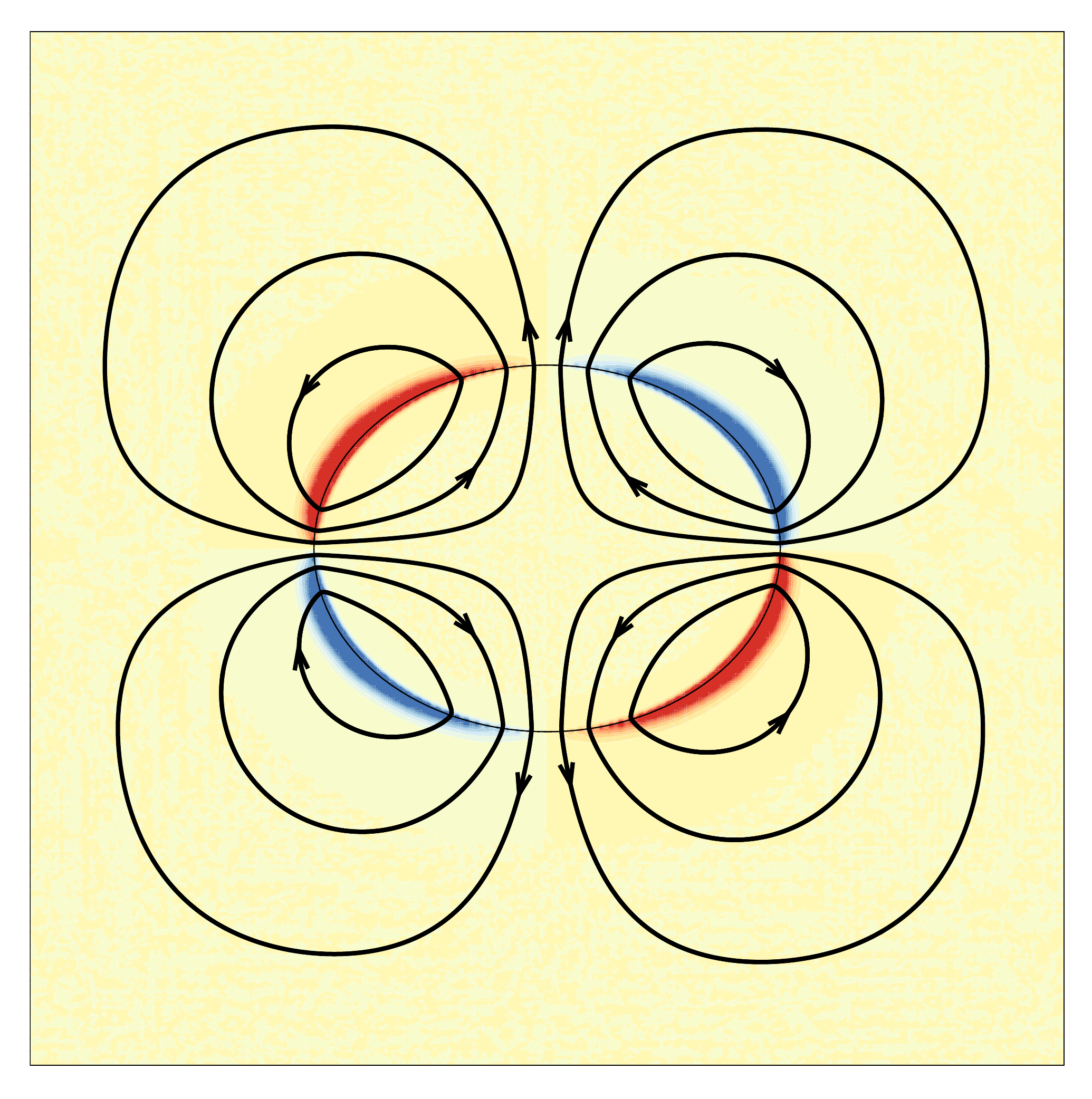}
\caption{\it Snapshot of scalar potential $\phi^o$, vector potential $\bm \psi^o$ and some trajectories for a time  $t = 10^{-3} \: s$   and a spatial approximation $n^2 = 512^2$ } 
\label{dropellips-phi}
\end{center}
\end{figure}

The simulations are performed on Cartesian meshes with degrees of freedom varying from $16^2$ to $512^2$; the initial ellipse is represented by a string of $128$, $256$ or $512$ markers. Figure (\ref{dropellips-phi}) shows the fields of scalar potential $\phi^o$ and instantaneous vector potential $\bm \psi^o$ for a time close to the initial state.

\begin{figure}[!ht]
\begin{center}
\includegraphics[width=10.cm,height=7.cm]{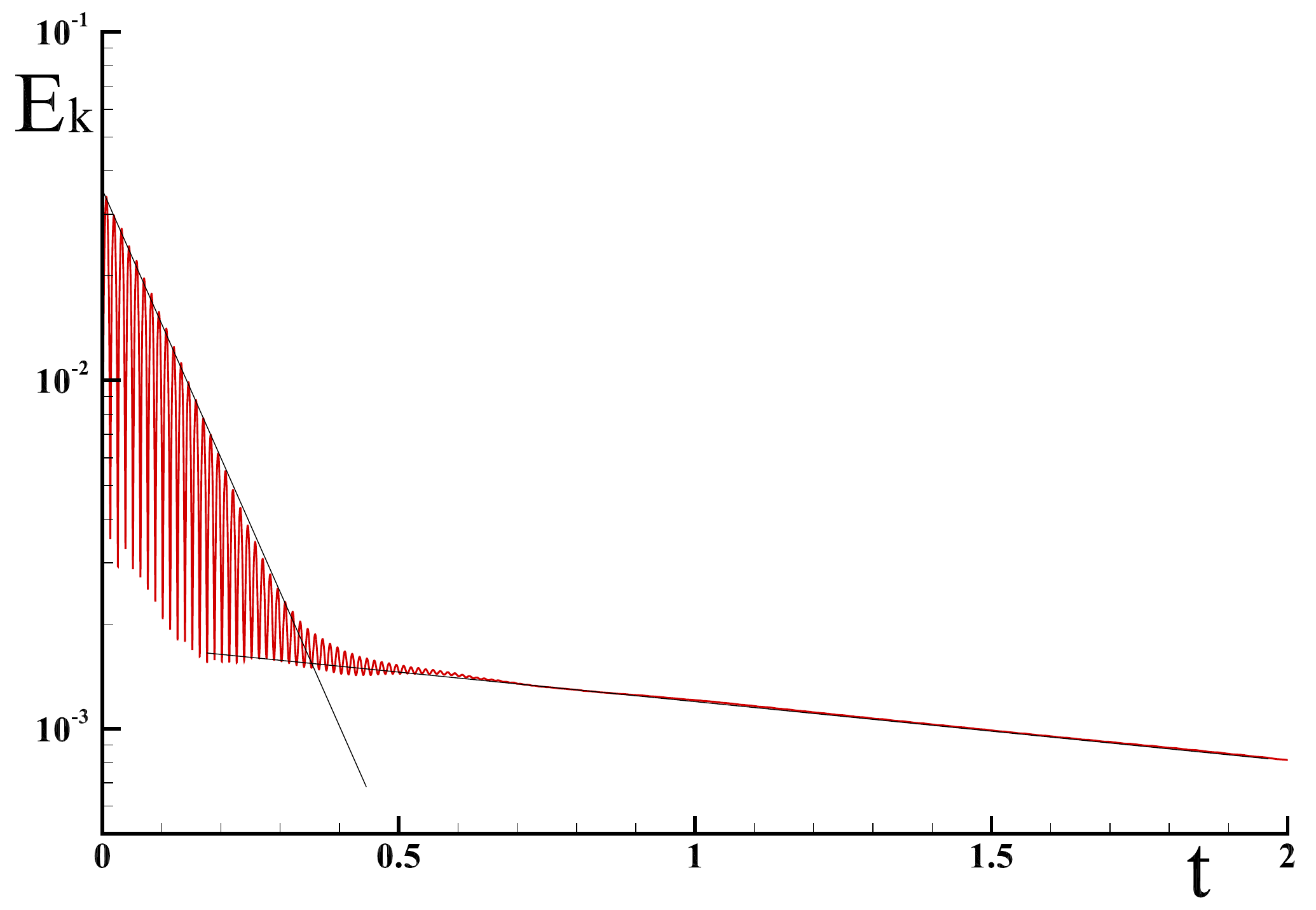}
\caption{\it Evolution of the kinetic energy $ E_k $ over time of an elliptical drop in a square containing gas for $\sigma = 2 \: 10^{-4} \: m^3 \: s^{-2}$, $\rho = 10^3 \: kg \: m^{-3}$, $dt = 10^{-4} \: s$ and  spatial approximation $n^2 = 64^2$. } 
\label{dropellips-a}
\end{center}
\end{figure}

The variations of the curvature within the source term ${\mathcal GRAD}  (\sigma \: \kappa \: \xi) $ generate an acceleration and a movement which tends to reduce the curvature, but the inertia leads to oscillations which slowly dampen given the low viscosity of water. This case test makes it possible to validate, among others, the specific nonlinear terms of the discrete formulation.
An example of the evolution of kinetic energy in the field is shown in figure (\ref{dropellips-a}).


We can clearly observe the two behaviors of the movement of the drop over time, first inertial and then viscous; the two lines in $E_k \propto exp(-\alpha \: t)$ represent their respective attenuations. For much greater times the velocity tends towards zero without parasitic currents and the difference in scalar potential takes the theoretical value $ \Delta \phi^o = 0.1$ to exactly satisfy the equality $- {\mathcal GRAD} \: \phi^o  + {\mathcal GRAD} \left(\sigma \: \kappa  \right) = 0$.

\begin{figure}[!ht]
\begin{center}
\includegraphics[width=10.5cm,height=7.cm]{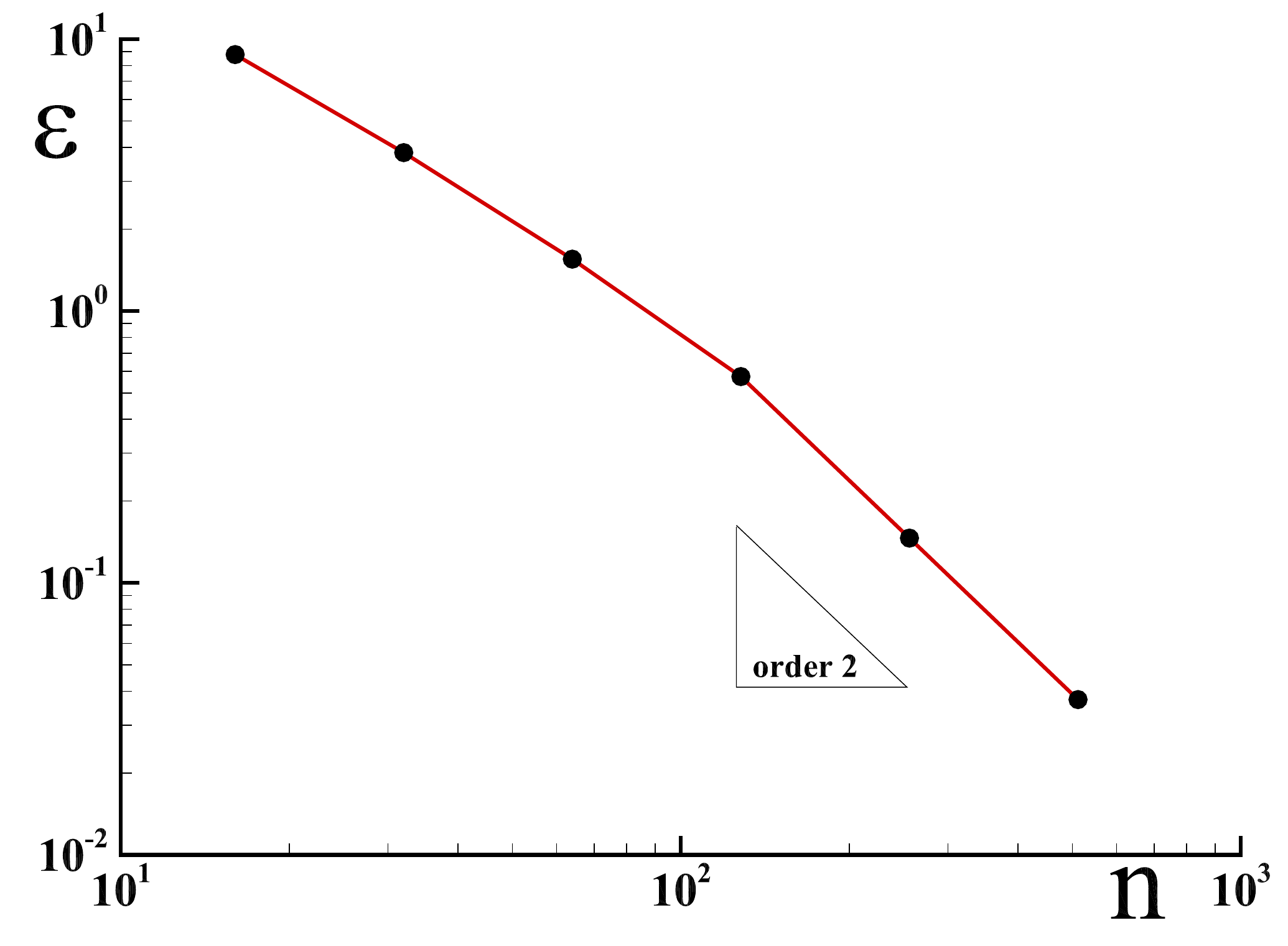}
\caption{\it Convergence of oscillation frequency $f$ with spatial approximation $n$. }
\label{convellips}
\end{center}
\end{figure}

The simulations carried out, with sufficiently weak time steps $(10^{-5} \: s)$  to neglect the errors in time and different spatial approximations, make it possible to calculate the rate of convergence of the numerical solution; this is given by table (\ref{frequence}) and figure (\ref{convellips}).
\begin{table}[!ht]
\begin{center}
\begin{tabular}{|c|c|c|c|c|c|c|c||c|}   \hline
 $ n^2$   & $16^2$  & $32^2$   &  $64^2$   & $128^2$   & $256^2 $ &  $512^2$   & $f_{\infty}$ & order \\ \hline  \hline
frequency& $31.928$ & $36.880$ & $39.154$  & $40.132$  & $40.560$ &  $40.669 $ & $40.706$ &  $1.97$  \\ \hline 
\end{tabular}
\caption{\it Frequency of oscillations $f$ with spatial approximation $n$. } 
\label{frequence}
\end{center}
\end{table}

The rate of convergence in space calculated from the value $f_{\infty} $, obtained using Richardson's extrapolation, is close to $2$; the theoretical value (\ref{frequency}) for the mode for $m = 2$ gives the value of $f_ {th} = 38.73 \: s^{-1}$. This difference can be explained by the nature of the problem treated, a little different from the non-viscous theory, the marked nonlinear effects, the confinement due to the presence of walls, and of course the numerical errors of discretization of the interface.

This example is not a complete study of the free oscillations of drops, but it does serve to verify that the numerical methodology presented simulates the phenomenon with an acceptable precision.
However, it is possible to affirm that the initial objective has been reached: multiple errors due to the processing of jump conditions, to the transport of interfaces by a front-tracking method, to the calculation of curvatures, to the compressible formulation, etc., are lower than the precision of the discrete formulation itself.

\textcolor{blue}{\subsection{Bubble of air rising in mercury}}

In order to give an idea of the robustness of the formulation, an example of two-phase flow is performed from the real values of the physical properties, density, viscosity and surface tension which vary significantly, spatially and temporally.
\begin{figure}[!ht]
\begin{center}
\includegraphics[width=3.5cm]{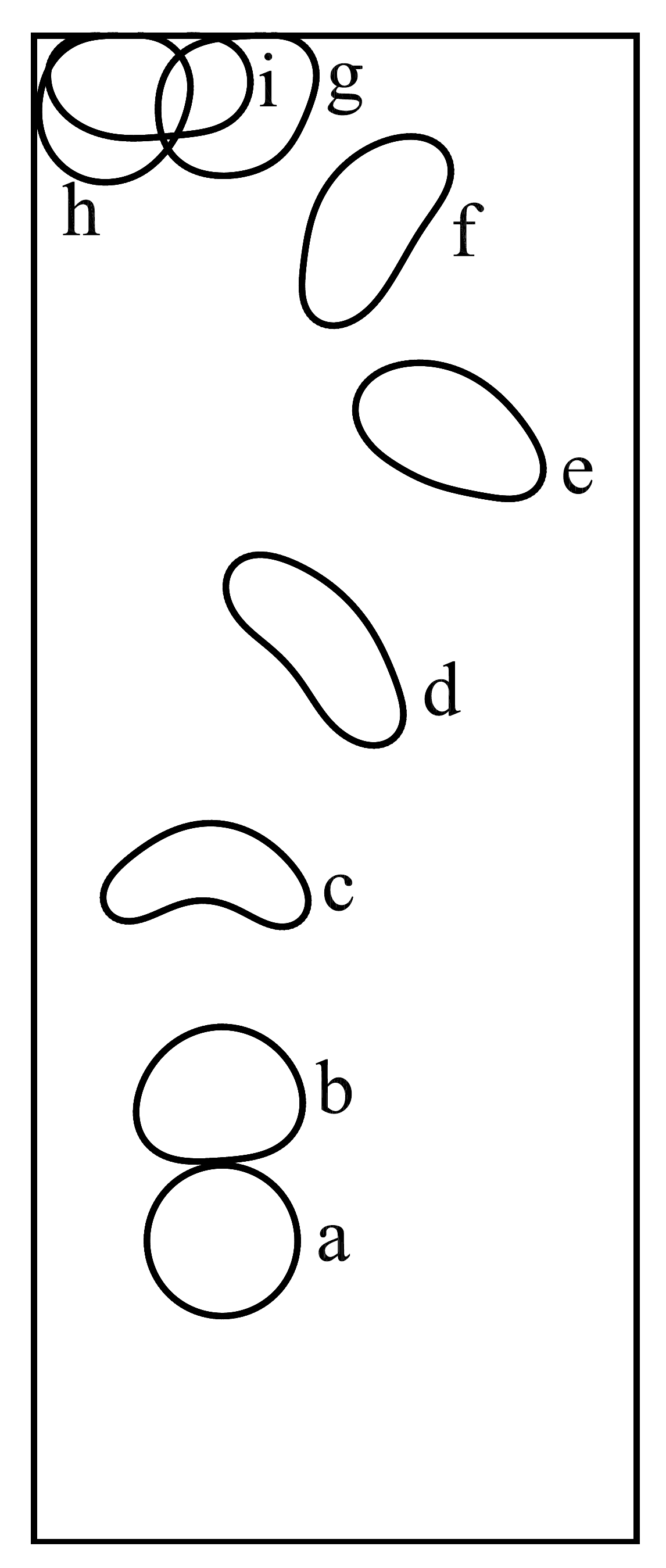}
\includegraphics[width=8.cm]{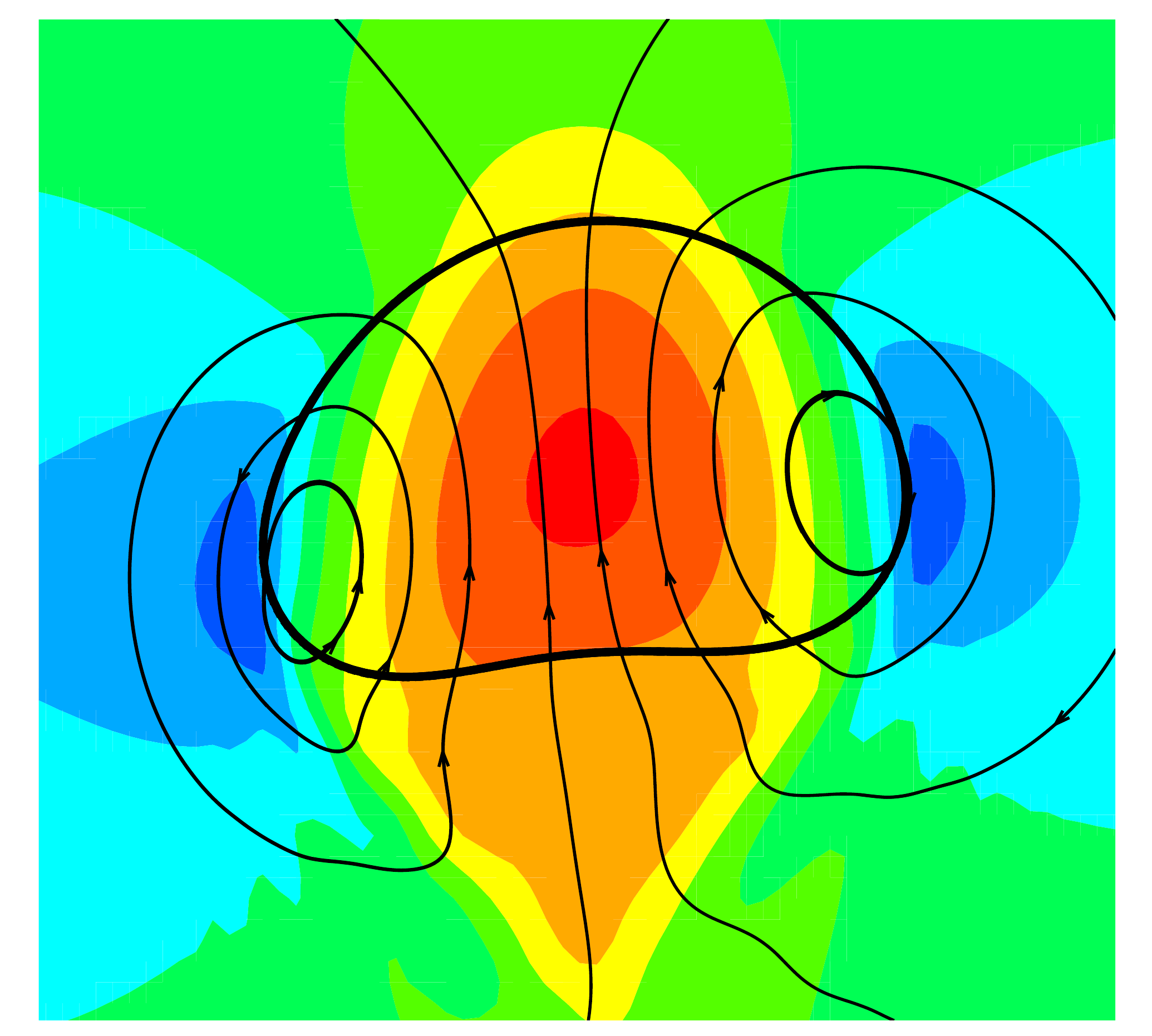} \\ 
\hspace{-25.mm} (a) \hspace{50.mm} (b)
\caption{\it (a) Shape of the bubble at different times: $a : 0 \: s$, $b : 0.5 \: s$, $c : 1 \: s$, $d : 1.5 \: s$, $e : 2 \: s$, $f : 2.5 \: s$, $g : 3 \: s$, $h : 3.5 \: s$, $i : 8 \: s$; (b)  snapshot of the vertical component of velocity at $t = 0.7 \: s$, trajectories and shape of the bubble extracted from the position of the $128$ markers.}
\label{cuvacier}
\end{center}
\end{figure}

A closed cavity of height $H = 0.1 \: m$ and width $L = 0.04 \: m$ is filled with mercury; an air bubble of radius $R = 0.01 \: m$ is initially positioned in $(x_0 = 0.0125, y_0 = 0.02)$ within the mercury and the whole is subjected to vertical gravity of $\mathbf g = -10 \cdot \mathbf e_y \: m \: s^{-2}$.
The properties of mercury (in SI units) are as follows: dynamic viscosity $\mu = 0.001526$, density $\rho = 13600$, longitudinal celerity $c = 1407$. Those of air are dynamic viscosity $\mu = 1.85 \: 10^{-5}$, density (constant) $\rho = 1.1768$, longitudinal celerity $c = 293$. The surface tension between the two fluids is taken as $\gamma = 0.5$. The boundary conditions of the cavity are adherent $\mathbf V = 0$.
The values are given for information only and may vary according to the chosen methodologies. The time step is $dt = 10^{-3} \: s$ and the total time of simulation is $8 \: s$. The regular Cartesian mesh adopted is $128 \cdot  256$ cells.
The air bubble is initialized by a string of $128$ markers defining the radius circle $R$. The position of the interface on $\Gamma$ segments is given by fast routines of differential geometry; it allows precise calculation of the density on each segment. The curvature is calculated from the dihedral angle between three consecutive markers.

The evolution of the shape of the bubble in time is shown by figure (\ref{cuvacier}a) at different times. 
Figure (\ref{cuvacier}b) shows a snapshot of the shape of the bubble at $t = 0.7 \: s$, some trajectories and the vertical component of the velocity field. The physical parameters lead to a divergence of less than $10^{-9}$ throughout the calculation.
The flow behind the bubble is highly inertial and unsteady. Even though the bubble motions weaken strongly after $3 \: s$ in the vicinity of the upper wall, the vortex flows persist for a very long time, given the kinematic viscosity of mercury $(\nu \approx 10^{-7} \: m^2 \: s^{-1})$. The field of the scalar potential is of order one and perfectly continuous; it allows a return to the pressure, only if necessary.


\textcolor{blue}{\section{Conclusions} }

The first part of this article presents the physical model, called discrete mechanics, and the numerical methodology which is closely associated with it. The general principles of the derivation of the law of motion abandon the notion of continuous medium to directly construct the discrete formal framework to establish a vector equation expressing the fact that the intrinsic acceleration of a material medium is equal to the accelerations applied to it in a given direction. This alternative formulation to the Navier-Stokes equations makes it possible to find the results of these with constant physical properties. The application of this new formulation to two-phase flows requires the establishment of conditions of jumps of the physical properties in line with the choice of expressing each physical effect in two contributions, one with curl-free and the second with divergence-free.

The formulation is mainly applied to very simple test cases where a theoretical solution of degree two is available. The resolution of these two-phase cases makes it possible to restore the theoretical solution with errors of the order of magnitude of the machine accuracy (without any numerical artifact), regardless of the chosen spatial approximation. The treatment of contact discontinuities does not affect the accuracy of the basic scheme. In the other cases \cite {Cal19a, Cal20a} and in general, this formulation is of order two in space and time. Its robustness can be evaluated on a flow with a high contrast of physical properties.

\vspace{5.mm}

\textcolor{blue}{\bf Author Contribution}

Author: Physical modeling, Conceptualization, Methodology, Research code,  Validation, Writing- Original draft preparation, Reviewing and Editing.

The paper has been checked by a proofreader of English origin.

\vspace{3.mm}

\textcolor{blue}{\bf Declaration of competing interest}

There are no conflict of interest in this work.

\vspace{3.mm}

\vspace{5.mm}
\bibliography{D:/calta/tex/database}



\end{document}